\documentclass[twocolumn]{aa} 
\usepackage{graphicx}
\usepackage[varg]{txfonts}
\usepackage{natbib}
\usepackage[version-1-compatibility]{siunitx}
\newcommand{\ramses}{{\tt RAMSES}}
\newcommand{\pluto}{{\tt PLUTO}}

\newcommand{\be}{\begin{equation}}
\newcommand{\ee}{\end{equation}}
\newcommand{\beal}{\begin{aligned}}
\newcommand{\eeal}{\end{aligned}}

\usepackage{hyperref}
\bibpunct{(}{)}{;}{a}{}{,} 
\usepackage{amsmath}
\usepackage{esdiff}
\usepackage{xcolor}
\usepackage[english]{babel}
\usepackage{multirow}

\usepackage{orcidlink}

\begin{document}

   \title{Disk fragmentation around a massive protostar: a comparison of two three-dimensional codes}

   \author{R. Mignon-Risse\,\orcidlink{0000-0002-3072-1496}
          \inst{1,2}
          \and
          A. Oliva\,\orcidlink{0000-0003-0124-1861} \inst{3,4}
          \and
          M. Gonz\'alez\,\orcidlink{0000-0002-3197-6095} \inst{1}
          \and
          R. Kuiper\,\orcidlink{0000-0003-2309-8963} \inst{5}
          \and
          B. Commer\c con\,\orcidlink{0000-0003-2407-1025} \inst{6}
          }

   \institute{Universit\'e Paris Cit\'e, Universit\'e Paris-Saclay, CEA, CNRS, AIM, F-91190 Gif-sur-Yvette, France \\
         \email{raphael.mignon-risse@apc.in2p3.fr}
         \and
         Universit\'e Paris Cit\'e, CNRS, AstroParticule et Cosmologie, F-75013, Paris
         \and
         Zentrum f\"ur Astronomie der Universit\"at Heidelberg, Institut f\"ur Theoretische Astrophysik, Albert-Ueberle-Stra{\ss}e 2, D-69120,
         Heidelberg, Germany
         \and
         Institute for Astronomy and Astrophysics, University of T\"ubingen, Auf der Morgenstelle 10, D-72076, T\"ubingen, Germany
         \and
         Faculty of Physics, University of Duisburg-Essen, Lotharstra{\ss}e 1, D-47057 Duisburg, Germany
         \and
         Univ Lyon, Ens de Lyon, Univ Lyon1, CNRS, Centre de Recherche Astrophysique de Lyon UMR5574, F-69007, Lyon, France
             }

   \date{Received ?; ?}
 
  \abstract
   {Most massive stars are located in multiple stellar systems. The modeling of disk fragmentation, a possible mechanism leading to stellar multiplicity, relies on parallel {3D} simulation codes whose agreement remains to be evaluated.
    }
   {Cartesian adaptive-mesh refinement (AMR) and spherical codes have been thoroughly used in the past decade to study massive star formation. 
   We aim to study how the details of collapse and disk fragmentation depend on those. 
   }
   {Using the Cartesian AMR code {\tt RAMSES} within its self-gravity-radiation-hydrodynamical framework, we compare disk fragmentation in a centrally-condensed protostellar system to the study of Oliva \& Kuiper (2020) performed on a grid in spherical coordinates using {\tt PLUTO}.
    }
   {To perform the code comparison, two {\tt RAMSES} runs are considered and give qualitatively distinct pictures. 
   On the one hand, when allowing for unlimited sink particle creation with no initial sink, Toomre instability and subsequent gas fragmentation leads to a multiple stellar system whose multiplicity is affected by the grid when triggering fragmentation and by numerically-assisted mergers.
   On the other hand, using a unique, central, fixed sink particle, a centrally-condensed system forms, similar to that reported in {\tt PLUTO}.
   Hence, the {\tt RAMSES}-{\tt PLUTO} comparison is performed with the latter: agreement between the two codes is found regarding the first rotationally-supported disk formation, the presence of an accretion shock onto it, the first fragmentation phase.
   Gaseous fragments form and their properties (i.e. number of fragments, mass and temperature) which are dictated by local thermodynamics, are in agreement between the two codes, considering that the system has entered a highly non-linear phase.
   Over the simulations, the stellar accretion rate is made of accretion bursts and continuous accretion, at the same order of magnitude.
   As a minor difference between both codes, fragments dynamics causes the disk structure to be sub-Keplerian in {\tt RAMSES} whereas it is found to be Keplerian and reaches quiescence in {\tt PLUTO}. We attribute this discrepancy to the central star being twice less massive in {\tt RAMSES} because of the different stellar accretion subgrid models in use rather than grid effects.
   }
   {In a centrally-condensed system, the agreement between {\tt RAMSES} and {\tt PLUTO} regarding many of the collapse properties and fragmentation process is good.
   On the opposite, fragmentation occurring in the innermost region and numerical choices (use of sink particles, grid) have a crucial impact {when similar but smooth initial conditions are employed} - more crucial than the code's choice - on the system's outcome, multiple or centrally-condensed.
   }


   \maketitle
%

\section{Introduction}
\label{sec:intro}

The multiplicity is higher for massive stars {($M>8\, \mathrm{M_\odot}$, where $\, \mathrm{M_\odot}$ denotes the solar mass)} than for their low-mass counterpart (see, e.g., \citealt{duchene_stellar_2013}), but the origin of this trend is {uncertain}.
Mechanisms producing multiple stellar systems include {dynamical interaction} {\citep{bate_formation_2002}}, pre-stellar core fragmentation {(\citealt{boss_fragmentation_1979}, \citealt{bonnell_fragmentation_1991}, \citealt{bonnell_fragmentation_1992}, \citealt{machida_collapse_2005}, \citealt{bate_stellar_2009}, \citealt{mignon-risse_collapse_2021-1})} and disk fragmentation {(\citealt{adams_eccentric_1989}, \citealt{shu_sling_1990}, \citealt{bonnell_massive_1994} in the case of a circumbinary disk, \citealt{kratter_fragmentation_2006}, \citealt{mayer_protoplanetary_2007}, \citealt{hennebelle_disk_2009}, \citealt{commercon_collapse_2011}, \citealt{wurster_disc_2019}, \citealt{oliva_modeling_2020}, \citealt{mignon-risse_collapse_2021-1})}.
The latter requires a scenario of disk-mediated accretion, which is currently supported by both observations (see e.g., \citealt{johnston_keplerian-like_2015}, \citealt{girart_circumestellar_2017}, and \citealt{sanna_discovery_2019} which also report a jet) and numerical experiments (e.g., \citealt{yorke_formation_2002} {\citealt{zinnecker_toward_2007}, \citealt{krumholz_formation_2009}}, \citealt{kuiper_circumventing_2010}, \citealt{kuiper_three-dimensional_2011} in the hydrodynamical case, \citealt{kolligan_jets_2018}, \citealt{mignon-risse_collapse_2021-1},
\citealt{mignon-risse_collapse_2021},
\citealt{commercon_discs_2022}, 
\citealt{oliva_modeling_2023-1},
\citealt{oliva_modeling_2023},
in the magnetic case). 
Observational constraints on disks around massive protostars are becoming increasingly numerous, and the Atacama Large Millimeter/submillimeter Array (ALMA) is now providing the first clues of disk fragmentation (e.g. \citealt{ilee_g1192061_2018}, \citealt{johnston_spiral_2020}).
Since the equations of hydrodynamics and the physics of fragmentation are highly non-linear, the questions of multiplicity (estimated by gaseous fragments or sink particles) and disk fragmentation are to be tackled in numerical simulations.
In particular, the {aspects relative to numerical methods, codes and their convergence are of main importance}.
In this paper, focus is made on the disk fragmentation origin of multiple stellar systems.

Sink particles have been introduced in numerical simulations in order to mimic the formation of stars {and their feedback} at smaller scales than what can be numerically resolved (\citealt{bate_modelling_1995} in smoothed particle hydrodynamics, SPH, \citealt{krumholz_embedding_2004} on grids, see \citealt{teyssier_numerical_2019} for a review).
Meanwhile, the use of sink particles may affect both the disk formation, equilibrium and fragmentation.
Hence, this topic is of main importance for the numerical studies of massive star formation and in line with the observational capabilities \citep{ahmadi_disc_2019}.
It has been recently studied in a work that focused on the disk properties and evolution in a radiation-hydrodynamical context, without using sink particles except for the central object (\citealt{oliva_modeling_2020}, hereafter OK20), that we refer to throughout this paper.
In particular, OK20 have shown that a numerical resolution of typically $20-30$~astronomical units (AU) in the disk was insufficient to resolve the disk spiral arms and subsequent fragmentation. On the one hand, introduction of sink particles under those circumstances is not physical because the fragmenting structures are not resolved. 
On the other hand, fragmentation, and possibly star formation, can be missed, and higher spatial resolution is required.
The introduction of sinks can be artificially enhanced because, unlike gas fragments, they can only be destroyed by merging with other sinks in most studies. For a finer description, mergers can even be forbidden, typically after the sinks reach the second Larson core mass \citep{rosen_unstable_2016}. 
This shows how (massive) stellar multiplicity, as predicted by computational studies, is dependent on numerical parameters.

{At a current epoch when many codes used for star formation incorporate self-gravity and radiation-(magneto-)hydrodynamics (see the review by \citealt{teyssier_numerical_2019}), and always more complex physics or chemistry, a code comparison is needed to identify numerical difficulties or caveats{ (as expressed in e.g. \citealt{klein_current_2006})}.
Some early studies compared grid-based codes {together} (see e.g. \citealt{bodenheimer_comparison_1979} and \citealt{boss_fragmentation_1979}) for 2D isothermal collapse with rotation, showing a good qualitative agreement{, and grid-based codes with SPH codes (\citealt{gingold_collapse_1981}, \citealt{bodenheimer_fragmentation_1981}, \citealt{gingold_reliability_1982}) in non-axisymmetric collapse with $m=2$ perturbations favoring binary formation, showing again a good qualitative agreement on the fragmentation process but also discrepancies about the further fate of the fragments (their coalescence or not)}.
More recent works mainly focused on comparing adaptive-mesh refinement (AMR) grid-based codes against SPH codes.
For instance, \cite{commercon_protostellar_2008} studied 3D isothermal collapse with rotation using \ramses{} and {\tt DRAGON} \citep{turner_binary_1995}, and 
\cite{federrath_modeling_2010} investigated stellar cluster formation with sink particles with an isothermal equation of state using {\tt FLASH}, \citep{fryxell_flash_2000}, and the SPH code developed by \citealt{bate_modelling_1995}).
They showed an encouraging agreement between both types of computational methods, provided that some resolution criteria associated with the Jeans mass are fullfilled.
Nevertheless, none of these studies focused on the problem of disk fragmentation in massive star formation nor solved the equations of radiation-hydrodynamics, going beyond the isothermal hypothesis.
}

{In this paper}, we use the \ramses{} code \citep{teyssier_cosmological_2002} with a radiation-hydrodynamical model {and self-gravity to address disk fragmentation in massive star formation, answering the need exposed above}.
We focus on a comparison with the two highest-resolution runs of OK20, performed with a modified version of \pluto{} \citep{mignone_pluto_2007}.

{Our numerical experiment is carried out assuming a massive, rotating, gravitationally-unstable pre-stellar core initially for simplicity, as widely done in the literature (e.g. \citealt{kuiper_solution_2014},
\citealt{rosen_unstable_2016}, \citealt{mignon-risse_new_2020}) and in the study of OK20; for reviews on the observational clues pointing to such structures, we refer the reader to \cite{beuther_formation_2007}, \cite{tan_massive_2014} and \cite{motte_high-mass_2018}.
The initial conditions are smooth, in the sense that no perturbation is applied to the density or velocity field.
For the core properties chosen here, a turbulent velocity dispersion of about ${\sim}0.5\, \mathrm{km \, s^{-1}}$ would be expected if the core follows the line-width-size relation of \cite{larson_turbulence_1981}. Nevertheless, the exact role of turbulence in driving fragmentation is not clear as the fragmentation process may depend on the spatial scale (see \citealt{kainulainen_high-fidelity_2013}), with thermal Jeans fragmentation at the sub-parsec scale in some cases despite $\mathrm{km/s}$ velocity dispersion (\citealt{beuther_fragmentation_2018}, \citealt{beuther_high-mass_2019}).
The main reason for the absence of turbulence in our models is to focus on the physics of disk fragmentation produced by the Toomre instability in a massive disk, keeping the setup as simple as possible.
}

{In the next section we present the numerical methods in the \ramses{} code.
As a preliminary step, presented in Sec.~\ref{sec:setup}, we identify a \ramses{} setup that can allow for a comparison of disk fragmentation around the central massive protostar with \pluto{}, and use this setup henceforth.
In Sec.~\ref{sec:num} we focus on the very early phases of the collapse until the first fragmentation era and the formation of a disk around the central protostar.
In Sec.~\ref{sec:disk} we study the disk evolution, fragmentation and the fragment properties.
In all sections, a comparison between the results offered by \ramses{} and \pluto{} is presented.
We conclude our study in Sec.~\ref{sec:ccl}.}
\\

\section{Methods}
\label{sec:model}

\subsection{Radiation-hydrodynamical model}

We use the \ramses{} code (\citealt{teyssier_cosmological_2002}, \citealt{fromang_high_2006}) to perform the following simulations.
\ramses{} is an AMR code which integrates the equations of radiation-hydrodynamics {and self-gravity}.
Radiative transfer is modeled with an hybrid radiative transfer method (\citealt{mignon-risse_new_2020}, akin to \citealt{kuiper_fast_2010}, \citealt{kuiper_makemake_2020}): we use the moment 1 (M1) method (\citealt{levermore_relating_1984}, \citealt{rosdahl_ramses-rt:_2013}, \citealt{rosdahl_scheme_2015}) to follow the propagation and absorption of radiation emitted by the primary star, and we use the flux-limited diffusion (FLD, \citealt{levermore_flux-limited_1981}, \citealt{commercon_radiation_2011}, \citealt{commercon_fast_2014}) otherwise.
{The M1 method explicitly solves the equations of conservation of the radiative energy and flux (i.e. a two-moment methods, in opposition to the one-moment FLD method), using the so-called "moment 1" closure relation \citep{levermore_relating_1984} giving the radiative pressure tensor as a function of the radiative energy and flux.
This closure relation ensures a correct behaviour in both the free-streaming limit and in the diffusion limit.
By evolving the radiative flux, in addition to the radiative energy, the directionality of a radiation beam and the associated shielding effects are better modeled than in the FLD method (\citealt{gonzalez_heracles:_2007}, \citealt{mignon-risse_new_2020}).}

We solve the following set of equations
   \begin{equation}
   \begin{aligned}
   \diffp{\rho}{t} + \nabla \cdot [\rho \boldsymbol{u}] 
   &= 0, \\
   \diffp{\rho \boldsymbol{u}}{t} + \nabla \cdot [\rho \boldsymbol{u} \otimes \boldsymbol{u} + P \mathbb{I}]
   &= - \lambda \nabla E_\mathrm{fld} + \frac{\kappa_\mathrm{P,\star} \rho}{\mathrm{c}} \boldsymbol{F}_\mathrm{M1} - \rho \nabla \phi, \\
   \diffp{E_\mathrm{T}}{t} + \nabla \cdot \biggl[\boldsymbol{u} \left( E_\mathrm{T} + P \right) \biggr]
   &= - \mathbb{P}_\mathrm{fld} \nabla : \boldsymbol{u} + \kappa_\mathrm{P,\star} \, \rho \mathrm{c} E_\mathrm{M1}  - \lambda \boldsymbol{u} \nabla E_\mathrm{fld} \\
   & \, \, \, \, \, \,+ \nabla \cdot \biggl[ \frac{\mathrm{c} \lambda}{\rho \kappa_{\mathrm{R,fld}}}\nabla E_\mathrm{fld} \biggr] - \rho \boldsymbol{u} \cdot \nabla \phi, \\
   \diffp{E_{\mathrm{M1}}}{t}  + \nabla \cdot \boldsymbol{F}_\mathrm{M1}
   &= - \kappa_\mathrm{P,\star} \, \rho \mathrm{c} E_\mathrm{M1} + \dot{E}_\mathrm{M1}^\star, \\
   \diffp{\boldsymbol{F}_\mathrm{M1}}{t}  + \mathrm{c}^2 \nabla \cdot \mathbb{P}_\mathrm{M1}
   &= - \kappa_\mathrm{P,\star} \, \rho \mathrm{c} \boldsymbol{F}_\mathrm{M1}, \\
  \diffp{E_{\mathrm{fld}}}{t} 
  +  \nabla \cdot [\boldsymbol{u}  E_\mathrm{fld}]
  &=
    - \mathbb{P}_\mathrm{fld} \nabla : \boldsymbol{u} 
    + \nabla \cdot \left( \frac{\mathrm{c} \lambda}
   	{\rho \kappa_{\mathrm{R,fld}}} \nabla E_{\mathrm{fld}} \right)\\
    & \, \, \, \, \, \, +\kappa_{\mathrm{P,fld}} \, \rho \mathrm{c} \left( \mathrm{a_R} T^4 - E_{\mathrm{fld}} \right), \\
   \Delta \phi &= 4 \pi \mathrm{G} \rho,
   \end{aligned}
   \end{equation}
   where, $\rho$ is the gas density, $\boldsymbol{u}$ is the velocity vector, $P$ is the gas thermal pressure, $\lambda$ is the flux-limiter in the FLD module, $\kappa_\mathrm{P,\star}$ is the Planck mean opacity computed at the effective temperature of the primary star, $\mathrm{c}$ is the speed of light, $\boldsymbol{F}_\mathrm{M1}$ is the M1 radiative flux, $\phi$ is the gravitational potential, $E_\mathrm{T}$ is the total energy which is defined as $E_\mathrm{T} = \rho \epsilon + 1/2 \rho u^2 + E_\mathrm{fld}$ (where $\epsilon$ is the specific internal energy), $E_\mathrm{M1}$ is the M1 radiative energy, $\mathbb{P}_\mathrm{fld}$ is the FLD radiative pressure, $\kappa_\mathrm{P,fld}$ is the Planck mean opacity in the FLD module (computed at the local gas temperature), $\kappa_\mathrm{R,fld}$ is the Rosseland mean opacity, $\mathrm{a_R}$ is the radiation constant, $\mathbb{P}_\mathrm{M1}$ is the M1 radiative pressure and $\dot{E}_\mathrm{M1}^\star$ is the injection term of the primary stellar radiation into the M1 module.

The term $\kappa_\mathrm{P,\star} \rho  \mathrm{c} E_\mathrm{M1}$ is the coupling term between the M1 and the FLD modules via the equation of temporal evolution of the internal energy, which is
   
   \begin{equation}
    C_\mathrm{v} \diffp{T}{t} 
   = \kappa_\mathrm{P,\star} \, \rho \mathrm{c} E_\mathrm{M1}
   + \kappa_{\mathrm{P,fld}} \, \rho \mathrm{c} \left(E_{\mathrm{fld}} - \mathrm{a_R} T^4  \right).
   \end{equation}
We employ the ideal gas relation for the internal specific energy $\rho \epsilon = C_\mathrm{v} T$,
where $C_\mathrm{v}$ is the heat capacity at constant volume.
\newline

{We have presented the radiation-hydrodynamical model incorporated in the \ramses{} code.
As we are interested in comparing \ramses{} and \pluto{}, we present in the following subsection their differences relevant to the formation of massive multiple stellar systems.}

\subsection{On the specifics of \pluto{} and \ramses{}}

Let us first present the numerical tools used in OK20, and how we can provide a complementary point of view.
\pluto{} (\citealt{mignone_pluto_2007}, \citealt{mignone_pluto_2012}) integrates the equations of hydrodynamics. 
Additionally, the equations for radiation transport and self-gravity are solved (see \citealt{kuiper_circumventing_2010} and \citealt{kuiper_makemake_2020} for details). 
Radiation transport is solved by considering frequency-dependent stellar irradiation via ray-tracing \citep{kuiper_fast_2010} and diffuse emission with the FLD method.
The spatial grid in OK20 uses spherical coordinates, centered on the (massive) protostar which would form via first and second hydrostatic core stages \citep{larson_numerical_1969}.
Gaseous fragments, representing potential companions and resolved down to first hydrostatic core scales, are not treated the same way as the central protostar. 
This way, their hydrodynamical properties can be followed and used to estimate whether they may form stellar companions.

The spherical grid is more adapted to simulating circumstellar disks than Cartesian grids, in particular for angular momentum conservation, thanks to the cell shape.
In case of fragment formation in the spherical grid, the cells around the fragments do not permit an angular momentum conservation (computed with respect to the fragment's center) for secondary disk formation as good as for the primary disk formation around the central object.
It is not clear though how angular momentum conservation compares, quantitatively, in Cartesian and spherical grids around those fragments.
However, it depends on spatial resolution, which is not uniform in OK20 nor in the present study.

The spherical grid employed in OK20 allows for a logarithmic spacing along the radial direction.
This leads to a particularly high spatial resolution in the disk inner regions, as compared to Cartesian AMR codes with the same total number of cells, and facilitates the implementation of ray-tracing techniques for the treatment of irradiation.
Indeed, the numerically fast ray-tracing has to occur along the first radial coordinate axis. 
This implies that the star should remain at the origin of the coordinate system. 
This does not imply that the star is fixed in space or is fixed with respect to the fragmented disk: by solving the equations in a frame co-moving with the primary star one allow the star to move with respect to the disk, while the gas in the computational domain feels additional forces from the co-moving grid{; such a co-moving grid has e.g.~been used in \cite{hosokawa_formation_2016}
with the same modified version of the \pluto{} code as OK20}; a comparison of the two different approaches is e.g.~given in \cite{meyer_burst_2019}.
Moreover, the resolution decreases with the distance to the primary star and some components of interest cannot be fully resolved, in particular in the outer parts of the disk or the large-scale cloud.

The study of OK20 shows that the finest resolution in numerical studies performed with Cartesian AMR codes is not always sufficient to get the converged Jeans length {(the one obtained by the highest-resolution runs)} sampled by several cells.
This can lead to spurious fragmentation and excessive formation of sink particles, or suppress fragmentation, depending on the sink formation algorithm. 
More quantitatively, they find that a finest resolution of $5$~AU is required to sample the converged Jeans length by several cells, in this particular setup.
The spatial scales on which fragmentation could occur (width of spiral arms, filaments, disk thickness) have to be resolved as well.
In the case of the disk pressure scale-height, this condition appears to be much less restrictive, except very close to the star (typically less than $30$~AU from the central star).
Let us note that, in the high-mass regime, the forming first core (of typically $1$~AU, \citealt{bhandare_first_2018}, see also \citealt{vaytet_simulations_2012} in the low-mass regime) immediately transforms into a secondary hydrostatic core \citep{bhandare_birth_2020}.
Hence, for now, the formation of second Larson cores (such as \citealt{bhandare_birth_2020}) will remain unresolved in disk fragmentation simulations.

{Here we address the problem of cloud collapse and disk fragmentation while comparing the results obtained with \pluto{} and \ramses{}, with an emphasis on the original \ramses{} simulations performed here}.
The AMR framework allows us to have a finer resolution than OK20 at radii larger than ${\sim}1000$~AU from the primary star.
It also provides the same resolution and cell shape, hence numerical diffusion, around fragments as around the primary star.

\subsection{Initial conditions}
\label{sec:frag_ci}

We use similar initial conditions as OK20.
We start from a massive pre-stellar core of mass $M_\mathrm{c} = 200 \, \mathrm{M_\odot}$ and radius $R_\mathrm{c} = 20\,625 \, \mathrm{AU} = 0.1$~pc,
whose density profile follows
\be
\rho(r) = \rho_\mathrm{0}  \left(\frac{r}{r_\mathrm{0}}\right)^{-3/2},
\label{eq:rho}
\ee
where $\rho_\mathrm{0}=2.89 \times 10^{-14} \mathrm{\,g\, cm^{-3}}$ at $r_\mathrm{0}=30$~AU, which sets the domain inner boundary in OK20.
{The density profile introduces a singularity at the center, which is not a problem in the numerical model of OK20 because there is a central sink cell with initial mass equal to the integral of the density profile within it. 
In the Cartesian code \ramses{} the density maximum will be set by the finest resolution; because of this profile, the properties of the central region slightly change with resolution and its evolution along with it (see the convergence study in Appendix~\ref{app:cvg}).}
{The index of the density profile power-law is in the range of massive dense cores (see \citealt{motte_high-mass_2018}).}
The mass within the inner $30$~AU region is about $0.01\, \mathrm{M_\odot}$.
The density profile results in a core mean density $\Bar{\rho} = 3.25 \times 10^{-18} \mathrm{\,g \, cm^{-3}}$ and an approximate free-fall time ranging from 
\begin{equation}
    \tau_\mathrm{ff} = \sqrt{\frac{3 \pi}{32 \mathrm{G} \Bar{\rho_\mathrm{0}}}} \simeq \, 0.4\, \mathrm{kyr}
    \label{eq:tauff}
\end{equation}
 in the inner $30$~AU region towards $37$~kyr for the entire core. In the rest of the paper, the time is given as the absolute time, i.e. starting at $t=0$~kyr.
Runs are performed up to $20$~kyr when possible, comparable to the simulation time in OK20, which is about half the core free-fall time. 
OK20 chose the rotation profile {(which is around the $z-$axis in the \ramses{} setup)}
\be
\Omega (R) = \Omega_\mathrm{0}  \left( \frac{R}{10 \mathrm{\,AU}} \right)^{-3/4},
\label{eq:omega}
\ee
where $\Omega$ is the angular frequency, $R$ is the cylindrical radius, which produces a rotational-to-gravitational energy ratio independent of the radius of the cloud.
Here, $\Omega_\mathrm{0}=9.84 \times 10^{-11} \mathrm{s^{-1}}$, and producing initially a uniformly sub-Keplerian azimuthal velocity and resulting in a rotational-to-gravitational energy ratio of $5 \%$ at the core scale.
Thanks to the angular momentum that is initially available at the center of the cloud, an earlier and more massive disk is formed compared to a solid-body rotation profile with the same rotational-to-gravitational energy ratio (see \citealt{meyer_forming_2018}, \citealt{meyer_episodic_2019}).
As in OK20, we use outflow boundary conditions and a uniform initial temperature of $10$~K.
Hydrodynamical simulations are performed with the Lax-Friedrich solver as in \cite{mignon-risse_new_2020}, respectively. For comparison, OK20 use the HLLC solver. Hence, we made the choice of a more stable but more diffusive solver than OK20 for the hydrodynamical simulations.

\subsection{Resolution and sink particles}
\label{sec:resol1}

The coarse resolution is level $5$ (equivalent to a $32^3$ regular grid) and the finest resolution level is $15$, resulting in a physical resolution of $2.5$~AU (see Appendix~\ref{app:cvg} for a convergence study).
Cells are refined so that the Jeans length is resolved by $12$ cells (see \citealt{truelove_jeans_1997}).
The comparison to OK20 is performed against their runs labeled {\tt x16} and {\tt x8}, which correspond to their highest-resolution and their next-to-highest-resolution runs, respectively.
{In this paper, those runs will be labeled {\tt PLUTOx8} and {\tt PLUTOx16}, respectively, whenever both are mentioned, {\tt PLUTO} otherwise.}
Boundary conditions for the velocity are outflows at the inner and outer boundaries in the radial direction and zero gradient for the density.
The finest resolution is $0.74$~AU in run {\tt PLUTOx8} and $0.368$~AU in run {\tt PLUTOx16} at $30$~AU from the grid origin, with logarithmic spacing at larger radii.
Accordingly, the resolution is about $7$~AU in run {\tt PLUTOx8} and $3.5$~AU in run {\tt PLUTOx16} at $300$~AU. 

Sink particles can be introduced to mimic the presence of a protostellar object.
Their implementation in {\tt RAMSES} is described in \citealt{bleuler_towards_2014}.
Sinks only interact gravitationally with the surrounding gas, and a Plummer gravitational softening is used with softening radius equal to four times the finest resolution.
The sink accretion radius is also set to be four times the finest resolution, $10$~AU.
For comparison, the radius of the central sink cell in OK20 is $30$~AU.
Accretion onto the sink occurs if gas within the sink cells is above a given density threshold. This threshold depends on the run resolution, and we want its value to be consistent with OK20.
Hence, we set the density threshold to $1.2 \times 10^{-13} \mathrm{g\, cm^{-3}}$
in the low-resolution run with a sink radius of $40$~AU presented in the Appendix~\ref{app:cvg}, which is similar to the density in the innermost cell (i.e. at radius $30$~AU) in OK20.
Then, we rescale the density threshold following the resolution dependency $\varpropto \mathrm{dx}^{-15/8}$ given in Eq.~11 of \cite{hennebelle_what_2020} to set it in the fiducial run.
{Not more than $10\%$ of the gas above this density can be accreted at each time step.}

{In the following, we investigate how collapse, fragmentation and the accretion properties depend on the numerical code.
However, before doing so, one must address the use of sink particles to be done in the {\tt RAMSES} run, for the sake of comparison with its use in the {\tt PLUTO} runs presented in OK20.}

\begin{figure}
\centering
    \includegraphics[width=9cm]{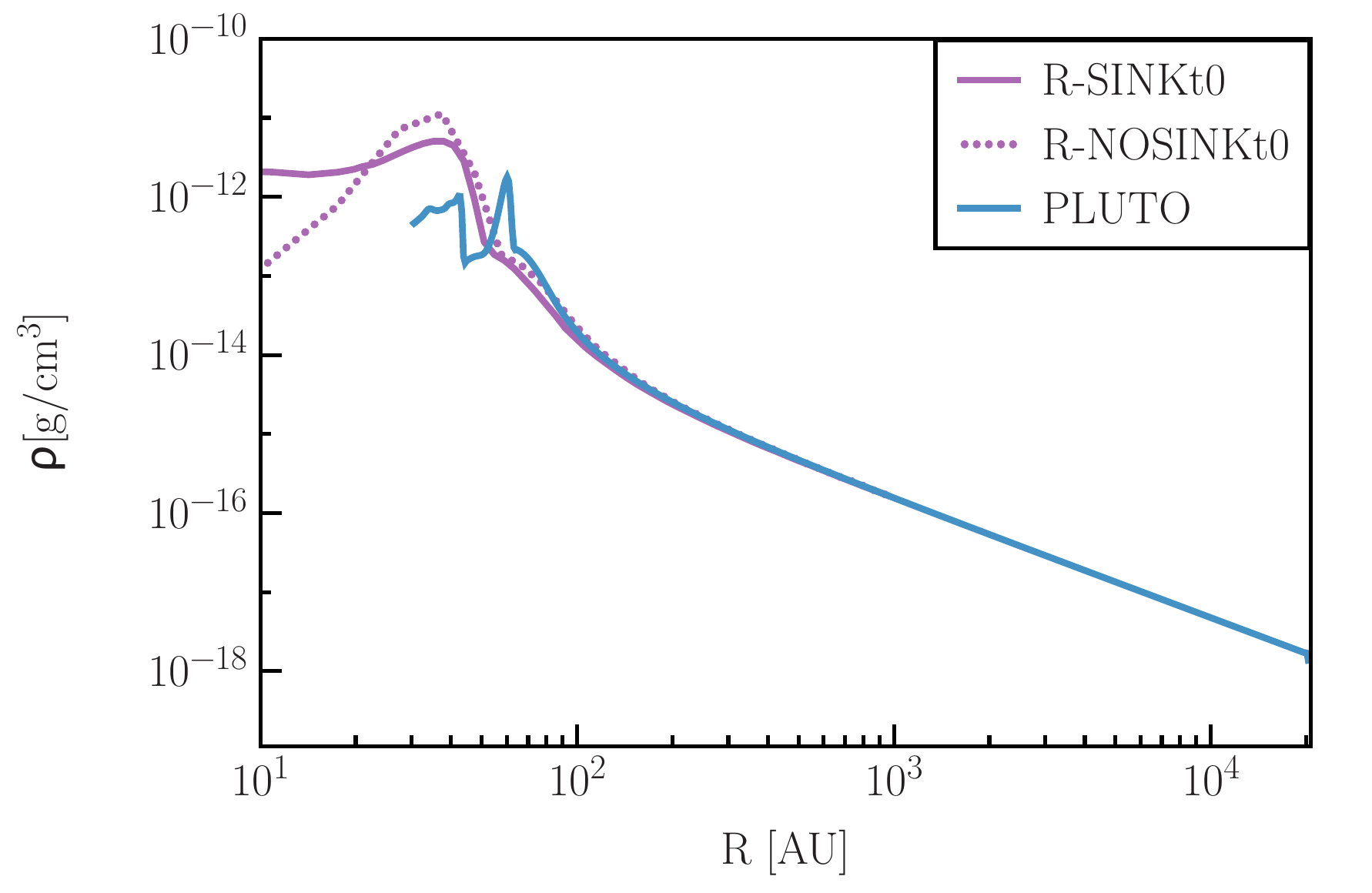} 
    \caption{Radial profile of the density in the $(x-y)-$plane at $t{\approx}2.5$~kyr in the {\tt R-SINKt0} (purple line), {\tt R-NOSINKt0} (purple, dotted line) and {\tt PLUTO} (blue line).
    At this stage, initial axisymmetry is maintained.
    Towards the center, the density settles in a plateau, forming a disk structure with an additional density bump in {\tt R-SINKt0} and {\tt PLUTO}, while it continuously decreases in {\tt R-NOSINKt0} and forms a ring around $30$~AU.}
    \label{fig:r_rho_25kyr}
\end{figure}

\section{Which setup for a {\tt RAMSES}-{\tt PLUTO} comparison? Centrally-condensed versus multiple system}
\label{sec:setup}

In order to perform the {\tt RAMSES}-{\tt PLUTO} comparison, we must make numerical choices regarding the {\tt RAMSES} setup, in particular on the use of sink particles to refine the scope of the present code comparison.
A first possibility is to use similar setups as already presented in previous {\tt RAMSES} projects (e.g. \citealt{mignon-risse_new_2020}).
In those setups, no sink particle is initially present and the number of sink particles to form eventually is not limited.
The run we present here, using this setup, is dubbed {\tt R-NOSINKt0} for the remainder of this section.
A second possibility, labeled {\tt R-SINKt0}, is to mimic some sink properties of the {\tt PLUTO} runs we aim to compare our results to.
In this second case, a single sink particle of initial mass $0.01\, \mathrm{M_\odot}$ is kept fixed at the center of the cloud during the entire simulation (but please be aware that this is not a requirement of the spherical grid code).
{This also acts as a way to flatten the innermost region, since the initial density profile is a power-law.}
We also forbid the formation of other sink particles in this second case.
In the following, we explore those two avenues, {\tt R-NOSINKt0} and {\tt R-SINKt0}, to identify which of the two is better suited for a deeper comparison with the {\tt PLUTO} runs presented in OK20.

We report two major differences between the {\tt R-SINKt0} and {\tt R-NOSINKt0} runs, which justify our further use of {\tt R-SINKt0} for comparison with {\tt PLUTO}.
First of all, during the first kyr of evolution, a ring nearly emptied of material at the center forms in {\tt R-NOSINKt0} while a disk with a density bump on top of it forms in {\tt R-SINKt0} and {\tt PLUTO}.
This is illustrated by Fig.~\ref{fig:r_rho_25kyr}, which shows the radial profile of the density at $t{\approx}2.5$~kyr in the $(x-y)$-plane.
The formation process of the density bump in {\tt R-SINKt0} and the ring in {\tt R-NOSINKt0} is very similar, and both are Keplerian, but the low-density material inside the ring is strongly sub-Keplerian and of very low density.
This difference is attributable to the gravitational influence of the central sink and its ability to retain the accreted gas.
Overall, {\tt R-SINKt0} and {\tt PLUTO} exhibit a central disk of radius ${<}70$~AU, while {\tt R-NOSINKt0} does not.

\begin{figure}
\centering
    \includegraphics[width=7.8cm]{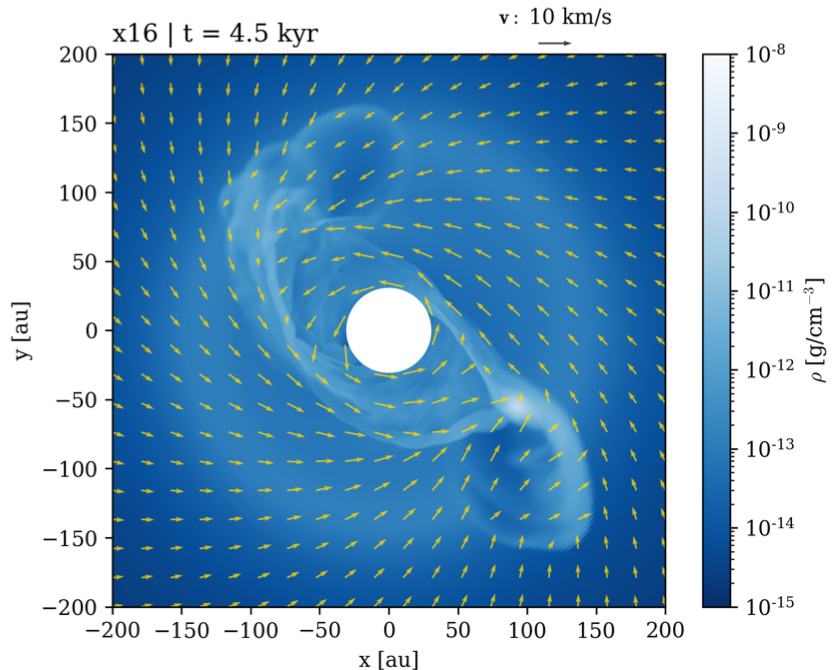}
    \includegraphics[width=8cm]{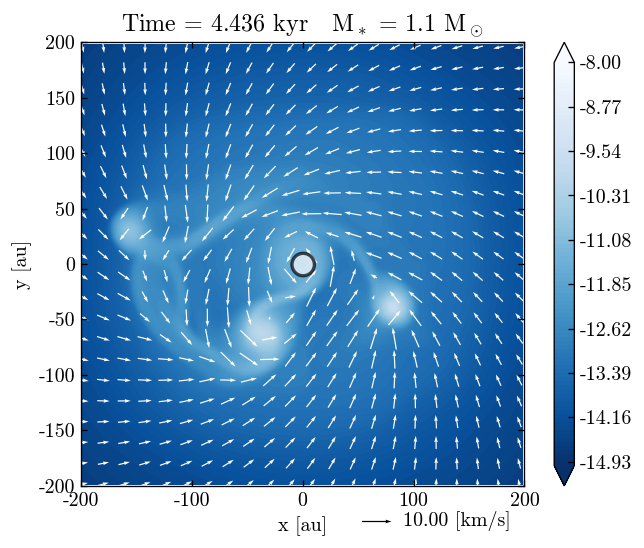}
    \includegraphics[width=8cm]{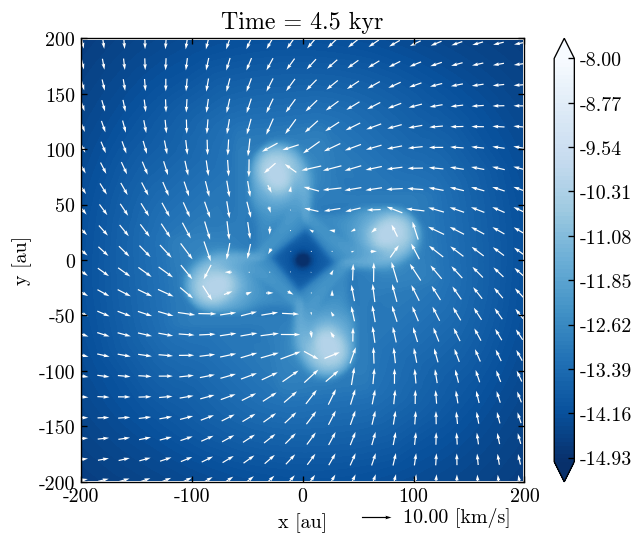}    
    \caption{Map of the density in logarithmic scale at the first fragmentation epoch ($t{\approx}4.5$~kyr) in the disk plane in {\tt PLUTOx16} (top), {\tt R-SINKt0} (middle) and {\tt R-NOSINKt0} (bottom).
    {\tt PLUTOx16} and {\tt R-SINKt0} form a centrally-condensed system whereas {\tt R-NOSINKt0} forms a multiple system.
    For the rest of the paper, we focus on a comparison between {\tt PLUTO} and {\tt R-SINKt0}.}
    \label{fig:fragmentation}
\end{figure}

Second, the {\tt R-SINKt0} and {\tt R-NOSINKt0} runs eventually form distinct systems: a centrally-condensed stellar system in {\tt R-SINKt0} and a multiple system in {\tt R-NOSINKt0}.
The fragmentation epoch producing this difference is illustrated in Fig.~\ref{fig:fragmentation} at $t=4.5$~kyr,  plotted together with the {\tt PLUTO} run.
In {\tt R-NOSINKt0}, fragmentation occurs on top of the Toomre-unstable ring, and appears to be triggered by the Cartesian grid.
Further in time, those fragments form sink particles that will merge until they form a binary system.
This mode of fragmentation and numerically-assisted mergers make uncertain the robustness of the final system's multiplicity.
In {\tt R-SINKt0} and {\tt PLUTO}, the first fragments are formed through the Toomre instability of the disk at the location of the density bump.
Those fragments end up being accreted by the central object after variable amounts of time.
Overall, the {\tt R-SINKt0} and {\tt PLUTO} runs give a similar qualitative picture, that of a centrally-condensed system, while the {\tt R-SINKt0} run forms a multiple system.

Parts of those differences between {\tt R-NOSINKt0} and {\tt R-SINKt0} are attributable to numerical methods and show the difficulties to model stellar multiplicity: promoting a centrally-condensed system on the one hand with a central sink particle, triggering preferential modes of fragmentation via the grid and influencing the stellar multiplicity through (numerical or physical?) sink mergers on the other hand.
The further evolution of those systems shows that, quite surprisingly, physical processes tend to conserve part of these initial differences, even though less than $7\%$ of the core free-fall time has elapsed at the time of fragmentation.
By now, it is unclear which of the two scenarios represents a more realistic model.
Very importantly, these results suggest that any fragmentation occurring at the center of such an idealized pre-stellar configuration is crucial in setting the final system's multiplicity, even though the global initial conditions are the same.
Further work on those aspects is needed but is beyond the scope of this paper.

As we aim to focus on disk fragmentation around a single object, investigating the central stellar mass growth, the modes of fragmentation, and the fragment properties, we choose the {\tt R-SINKt0} run for comparison with {\tt PLUTO}.
For the remainder of this paper, we refer to the {\tt R-SINKt0} run as the {\tt RAMSES} run, for conciseness.

\section{From cloud collapse to disk formation in a centrally-condensed stellar system}
\label{sec:num}

{In the following, we explore the early phases of massive star formation described in our numerical experiment, namely: first radial equilibrium reached, formation of a central accretion disk exhibiting a density bump, followed by the first fragmentation phase triggered by Toomre instability.
At each step, we compare the outcomes of the {\tt RAMSES} and the {\tt PLUTO} runs, while focusing our in-depth analysis on the {\tt RAMSES} simulations original to this paper.}

\subsection{{Equilibrium, first fragmentation era}}

\begin{figure}
\centering
    \includegraphics[width=8cm]{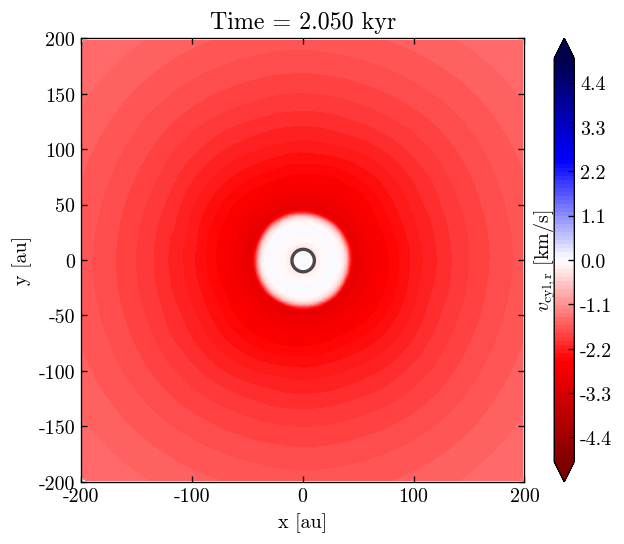} 
    \includegraphics[width=9.5cm]{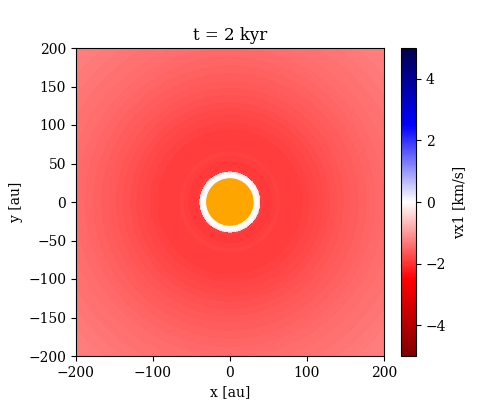}
    \caption{Map of the cylindrical radial velocity at $t{\approx}2$~kyr in the disk midplane in the {\tt RAMSES} run (top panel) and in the {\tt PLUTOx16} run (bottom panel).
    The sink particle in {\tt RAMSES} is represented by the white-filled, black circle; the sink cell in {\tt PLUTO} is represented by the orange circle.
    Gas is dominated nearly everywhere by infall motions (red regions) until it shocks onto the flattened core, resulting in a radial velocity close to zero.}
    \label{fig:accshock}
\end{figure}

The cloud is initially gravitationally-unstable and leads to a global infall motion.
The density in the central regions increases until it becomes optically-thick: it switches from isothermal to adiabatic.
This is the first Larson core.
As it contracts further, the gas temperature heats up and the thermal pressure increases accordingly.
At the border of the Larson core, the gradient of thermal pressure becomes strong enough to halt the infalling material.
Further in time, centrifugal acceleration increases due to infalling rotating material and finally dominates over thermal pressure gradients in setting a rotationally-supported structure: an accretion disk is born.
Figure~\ref{fig:accshock} shows the map of the radial velocity in the disk plane at ${\approx}2$~kyr in \ramses{} (top) and in \pluto{} (bottom).
The accretion shock onto the disk is visible as the sharp transition between the red, infall region, and the white, disk region.
\ramses{} and \pluto{} agree on the qualitative picture, namely the onset of the adiabatic stage, formation of a Keplerian disk and formation of an accretion shock.
They also agree on the size of the disk and its associated formation timescale.


After $2$~kyr, a density bump forms between $30$ and $50$~AU in the {\tt RAMSES} run, as shown on Fig.~\ref{fig:r_rho_25kyr} and mentioned in Sec.~\ref{sec:setup}.
This structure will be studied in more details in Sec.~\ref{sec:bump}.
In a qualitative view, rotating infalling material appears to have "bounced" onto the central region.
As mentioned in Sec.~\ref{sec:setup}, a similar structure is present in \pluto{} runs (see Fig.~\ref{fig:r_rho_25kyr}) but this structure is sharper in \pluto{} as it is caused by several axisymmetrical accretion shocks propagating through the disk.

At this stage, within the innermost $30$~AU, the total gas+sink mass is $0.37\, \mathrm{M_\odot}$ in {\tt RAMSES} against $0.47\, \mathrm{M_\odot}$ at the same time in {\tt PLUTO}. 

In {\tt RAMSES} as in \pluto{}, the bump is where the disk becomes (the most) Toomre-unstable and the first fragments form, breaking the axisymmetry, between $t=3$~kyr and $t=4$~kyr.
The location of the bump, when fragmenting, is between $30$ and $50$~AU.
The density map after fragmentation is displayed in Fig.~\ref{fig:fragmentation}.
This is the first fragmentation epoch. 
We note that the number of fragments is different between the codes: $2$ in {\tt PLUTO} (one is being sheared in the view of Fig.~\ref{fig:fragmentation}), $3$ in the {\tt RAMSES} run.
The initial perturbations are certainly introduced by numerical errors.
Otherwise, $m=4$ symmetry (following the Cartesian grid) in the \ramses{} run and axisymmetry in the \pluto{} runs should be perfectly conserved.
It can be seen that fragments moved from their initial radius (i.e. the radius of the fragmenting structure). Indeed, they evolve on eccentric orbits, interacting with the central sink, with the background disk and with the other fragments.
The disk size is similar in both codes.
At this stage, due to different sink algorithms, the mass is naturally distributed in a different manner between the sink and the gas in the two codes : the sink mass is $1.14\, \mathrm{M_\odot}$ in {\tt RAMSES} against $1.66\, \mathrm{M_\odot}$ in {\tt PLUTO}. 
Zooming-out, the total gas+sink mass is $2.22 \, \mathrm{M_\odot}$ within $100$~AU and $2.57\, \mathrm{M_\odot}$ within $200$~AU, in the {\tt RAMSES} run.
For comparison, the total gas+sink mass is $2.00 \, \mathrm{M_\odot}$ within $100$~AU and $2.56 \, \mathrm{M_\odot}$ within $200$~AU, in the {\tt PLUTO} run, which give relative differences between codes of $11\%$ and $<1\%$, respectively.

To sum up, we followed the very early phases of massive protostellar collapse. 
\ramses{} and \pluto{} runs agree on the formation of a rotationally-supported structure (disk) and on the presence of an accretion shock at its border.
A density bump is formed in both codes but for distinct reasons (see Sec.~\ref{sec:bump}).
It causes the first disk fragmentation phase as it is Toomre-unstable, at about the same time in \ramses{} and \pluto{} runs.

\subsection{{Density bump formation: an interplay between pressure gradient, centrifugal and gravitational accelerations}}
\label{sec:bump}

\begin{figure}
\centering
    \includegraphics[width=9cm]{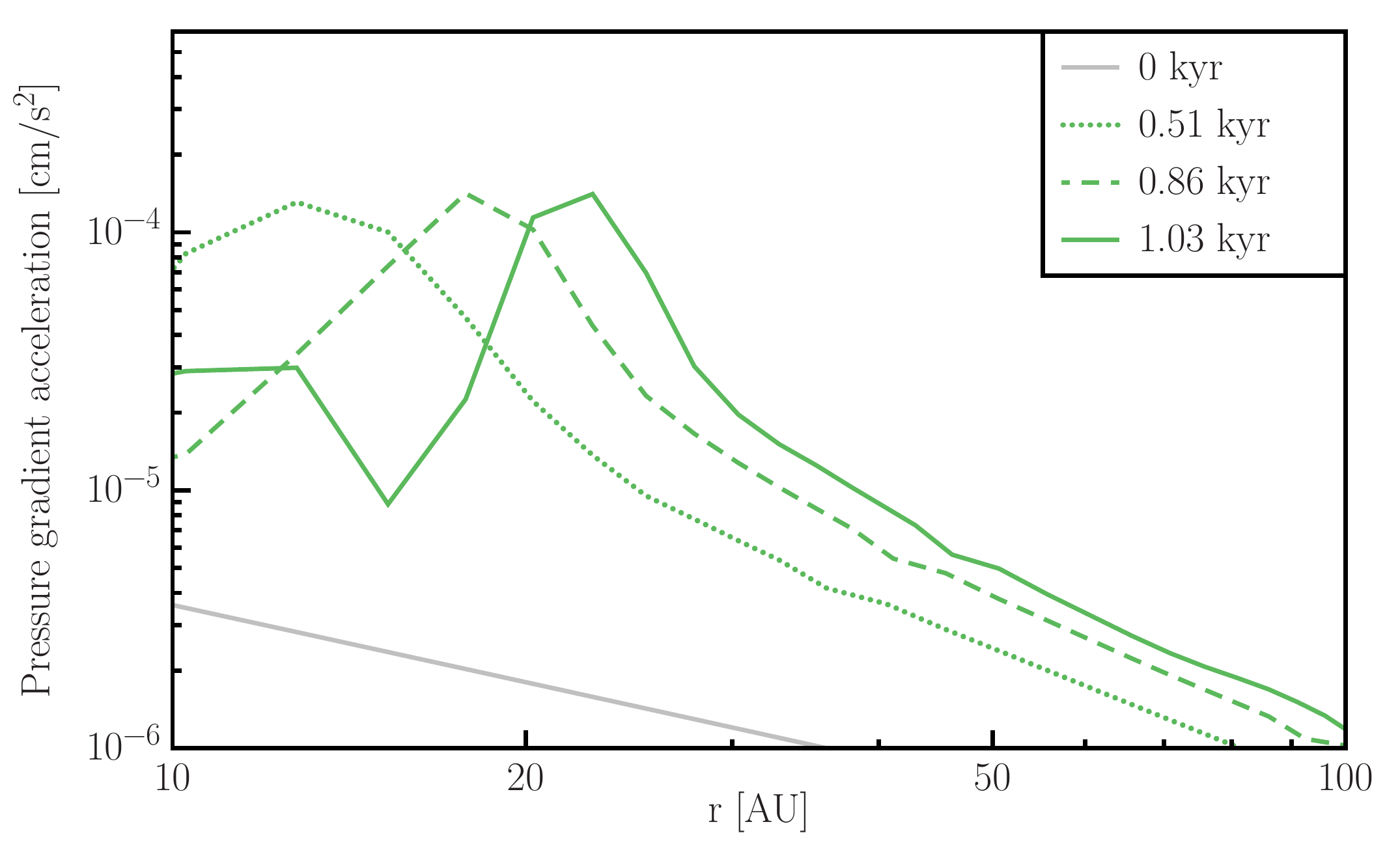} 
    \includegraphics[width=9cm]{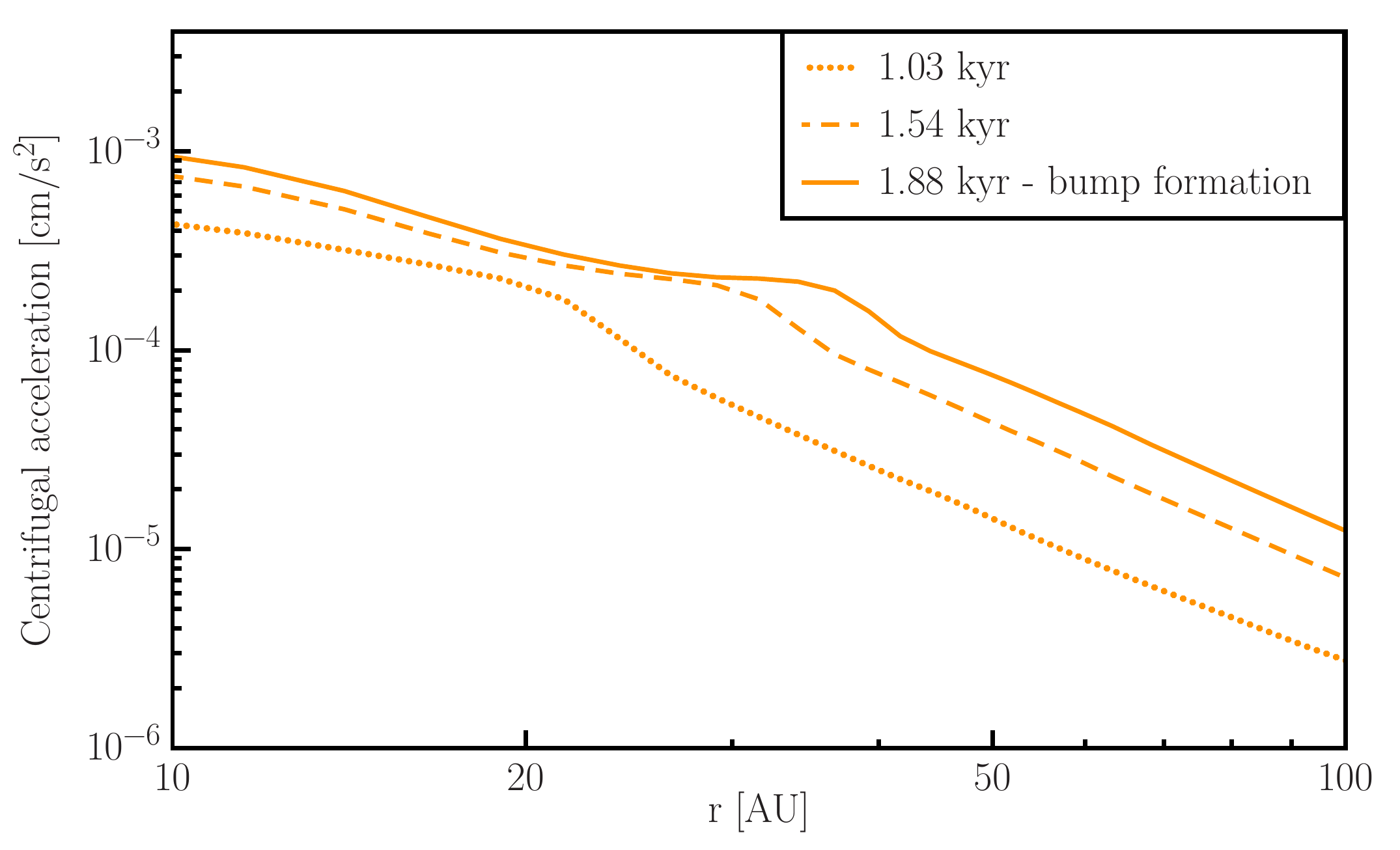}
    \includegraphics[width=9cm]{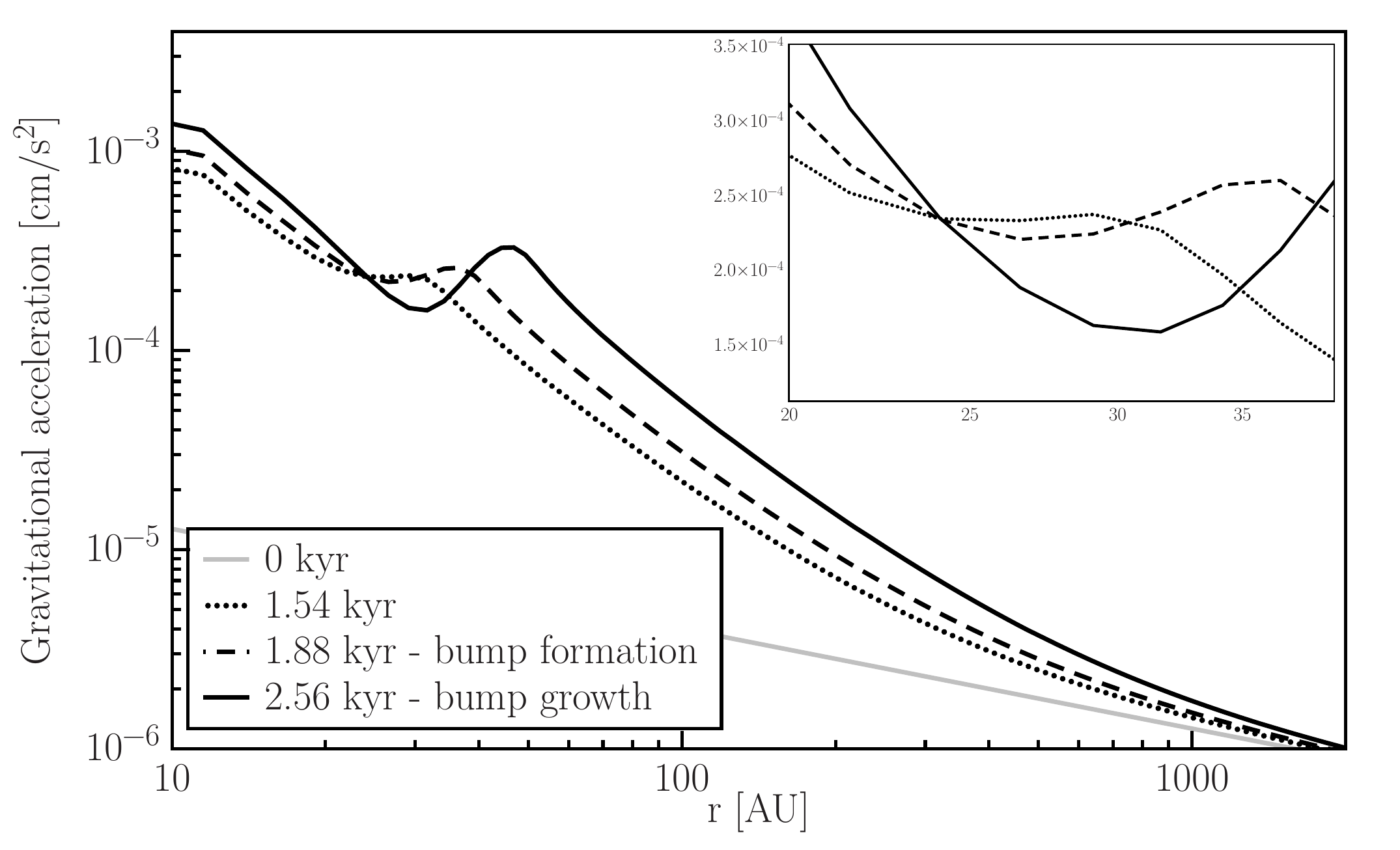}
    \caption{Radial (cylindrical) acceleration due to thermal pressure gradient (top panel), centrifugal acceleration (middle panel) and gravitational acceleration (bottom panel) at distinct epochs to study the density bump formation, which is responsible for the first fragmentation.
    Only the radial gravitational acceleration is directed inward.
    In the zooming-view of the bottom panel, it can be seen that the inward gravitational acceleration weakens around $30$~AU at $t{\lesssim}2$~kyr, so that gas accumulates into a bump. 
    }
    \label{fig:accs}
\end{figure}

We aim to understand how does the density bump form in the disk in {\tt RAMSES} as it further triggers the first fragmentation phase.
In order to do so, we compute the relevant accelerations at work in the cylindrical radial direction: centrifugal, thermal pressure gradient and gravitational accelerations.
Their radial profile is shown in Figure~\ref{fig:accs} for the epochs of interest depending on the acceleration at play.

First, we find that the sum of the outward thermal pressure gradient and centrifugal accelerations balance the inward gravitational acceleration, ensuring equilibrium.
This is linked to the pressure gradient stopping the gas at the border of the first hydrostatic core, indicated by the peak in the pressure gradient acceleration profile (top panel of Fig.~\ref{fig:accs}).
These forces are responsible for the accretion shock presented before.
Finally, this structure (future disk) expands as the peak shifts towards larger radius (see Fig.~\ref{fig:accshock}).
It is flattened in the vertical direction due to rotation (e.g. \citealt{black_evolution_1976}).

The infall of material brings additional specific angular momentum which eventually contributes to the centrifugal acceleration.
Indeed, the initial rotation profile gives $v_\phi \varpropto r_\mathrm{cyl}^{1/4}$.
Angular momentum conservation for a portion of gas implies that its cylindrical velocity at a given final radius $r_\mathrm{f}$ is $v_{\phi,\mathrm{f}}=v_{\phi,\mathrm{i}} r_\mathrm{cyl,i}/r_\mathrm{cyl,f}>v_{\phi,\mathrm{i}}$ (because of the infall motion).
While the view exposed here neglects mixing, it is clear that $v_\phi$ increases locally due to the input of specific angular momentum associated with the infall.
Since the centrifugal acceleration at a given cylindrical radius is proportional to $v_\phi^2$ and $v_\phi$ increases locally, as shown above, it increases very rapidly (see the bottom panel of Fig.~\ref{fig:accs}) until it becomes the dominant force counter-balancing the gravitational acceleration: we refer to this equilibrium structure as a disk.

Initially, the gravitational acceleration increases in the central region as the mass increases.
As the disk forms around the first Larson core, the mass distribution becomes highly anisotropic (while remaining axisymmetric) because of the centrifugal acceleration \citep{larson_collapse_1972}.
As a consequence of this new distribution, the gravitational acceleration decreases around $30$~AU, as shown in the zoomed view in the middle panel of Fig.~\ref{fig:accs}.
Indeed, when mass accumulates at the border of the disk, it reduces the inward gravitational acceleration at smaller radius (see Appendix~A of \citealt{tohline_ring_1980}).
The anisotropy is crucial: in a spherically-symmetric density distribution, a gas portion located at a given radius only feels the gravitational acceleration due to the mass enclosed within this radius; the isotropic density distribution located further out does not contribute to the gravitational acceleration.
In the same region, the centrifugal acceleration is not reduced because of angular momentum conservation: it can either stay constant, or increase because of additional input of infalling, rotating material with higher specific angular momentum.
Then, the centrifugal acceleration starts to dominate over the gravitational acceleration locally.
This is a runaway process: as the gas is given a positive radial velocity and is driven to larger radius, the gravitational acceleration is reduced even more, until an axisymmetric density bump forms.
Two mechanisms moderate or stop this process.
First, the accretion onto the sink removes material from the grid, which thus cannot participate to the bump growth.
The second mechanism is the fragmentation of the disk.
The disk is globally Toomre-unstable, but the location at which it is most unstable is in the density bump.
The bump growth stops when the disk fragments, at the bump location, which occurs on an orbital timescale \citep{norman_fragmentation_1978}.

We address the robustness of the bump formation process (or its somewhat equivalent ring structure when no sink is present initially) in Appendix~\ref{app:bump}.

\subsection{Disk growth}

\begin{figure}
\centering
    \includegraphics[width=9cm]{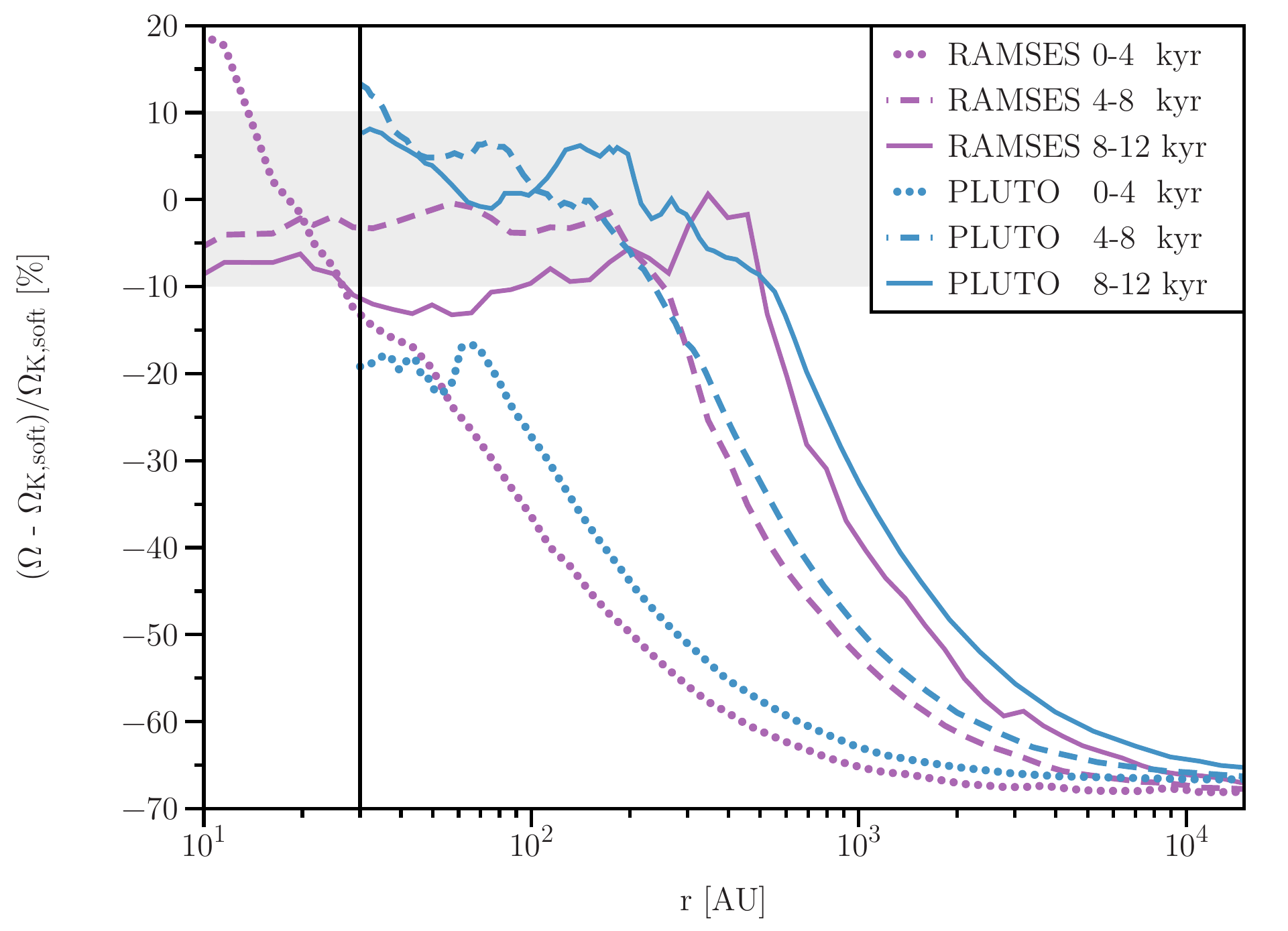} 
    \caption{{Radial profile of the deviation from the Keplerian frequency.
    Purples curves refer to the {\tt RAMSES} run and blue curves refer to the {\tt PLUTOx8} run.
    The Keplerian frequency is corrected by the sink softening length (Eq.~\ref{eq:col_fragrkss}). 
    For each radius, we compute the azimuthal median and average it in time (as indicated in the plot legend). The use of a median is meant to get rid of non-axisymmetries such as spiral arms (see OK20). 
    The gray area points to a deviation of $\pm10\%$ with respect to Keplerian frequency. The vertical line indicates the sink cell radius in {\tt PLUTO} runs.
    Until $t{\sim}12$~kyr, disks are comparably Keplerian in {\tt RAMSES} and {\tt PLUTOx8}}.}
    \label{fig:diskkepl1}
\end{figure}

{
Following the first fragmentation era and the formation of a dense halo (see Fig.~\ref{fig:fragmentation}), the central accretion disk grows as rotating gas falls in.
As a diagnostic of the disk structure (and in particular, its outer radius), we compute the deviation from Keplerian frequency of the rotating gas around the central sink.
We correct the Keplerian frequency by the sink "softening length" which softens the gravitational force \citep{bleuler_towards_2014}. Hence, the modified Keplerian angular frequency $\Omega_\mathrm{K,soft}$ is given by
\be
\Omega_\mathrm{K,soft} =\sqrt{\frac{\mathrm{G}M_\mathrm{gas}(<r)}{r^3} + \frac{\mathrm{G}M_\mathrm{sink}(r)}{(r^2+\mathrm{r_{soft}^2} )^{3/2}}},
\label{eq:col_fragrkss}
\ee
where $\mathrm{r_{soft}}$ is four times the finest spatial resolution here, $M_\mathrm{gas}({<r})$ is the mass enclosed in a radius $r$ and $M_\mathrm{sink}(r)$ is the sink mass for $r \ge \mathrm{r_{soft}}$ and the fraction of the sink mass enclosed in a radius $r$ for $r<\mathrm{r_{soft}}$.}

{Figure~\ref{fig:diskkepl1} shows the deviation with respect to Keplerian frequency from $0$ to $12$~kyr, averaged over $4$~kyr intervals, in the {\tt RAMSES} and {\tt PLUTO} runs.
The gray area represents a $\pm 10\%$ deviation with respect to perfect Keplerian frequency.
This Figure shows a region of gas compatible with Keplerian rotation (i.e. in centrifugal equilibrium), already from the $[0,4]$~kyr time interval, in the {\tt RAMSES} run.
This is the accretion disk around the central star growing with time.}

{We notice that the deviation from Keplerian frequency in the {\tt RAMSES} and {\tt PLUTO} runs rarely deviate by more than $10\%$ from the other, showing an overall correct agreement at following the disk formation epoch in terms of rotation support (deviation from Keplerian frequency) and disk size (transition from a mainly rotational support to infall motion).
The disk build-up appears slightly more rapid in the {\tt PLUTO} run, for any time interval.
This suggests a faster collapse in {\tt PLUTO} than in {\tt RAMSES}.
A possible explanation comes from the gas mass located outside the pre-stellar core in the {\tt RAMSES} run, which may slightly slow down the collapse, in a similar manner as reported in \cite{federrath_modeling_2010}.
The higher resolution achieved in the central regions in {\tt PLUTO} may also contribute to a faster collapse than in the {\tt RAMSES} run.
}

\section{Disk dynamical state: evolution and fragmentation}
\label{sec:disk}

\begin{figure}
\centering
    \includegraphics[width=9cm]{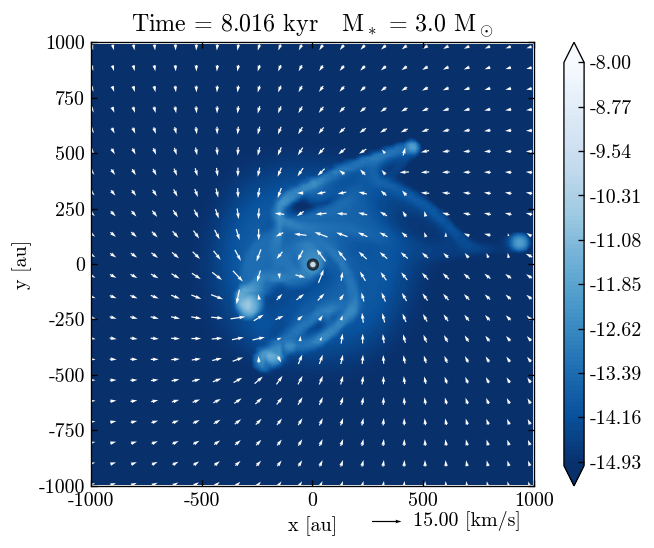}
    \includegraphics[width=9cm]{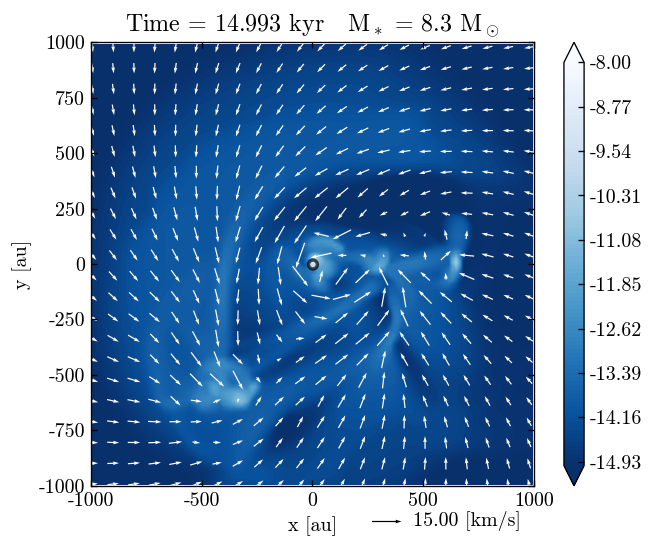} 
    \caption{Map of the density in logarithmic scale at $t{\approx}8$~kyr (top panel) and $t{\approx}15$~kyr (bottom panel), in the disk midplane in the \ramses{} run.}
    \label{fig:rhomap_dynamical}
\end{figure}

{In the following, we focus on the disk dynamical state and on the properties of its fragments, as those could collapse to form stellar companions.
Part of the time evolution of the system in \ramses{} is illustrated with density maps in Fig.~\ref{fig:rhomap_dynamical}.
Hence, we start by looking at the stellar accretion history because its gravitational influence is decisive for the gas dynamics which tends to settle in to a gravito-centrifugal equilibrium around the central star \citep{zinnecker_toward_2007} and for the disk stability as it sets the relative importance of the disk self-gravity \citep{kratter_gravitational_2016}.
This is also an opportunity to see how different subgrid methods for accretion compare together.
}

\subsection{Mass accretion history of the central star}
\label{sec:msinkok20}

\begin{figure*}
\centering
    \includegraphics[width=9cm]{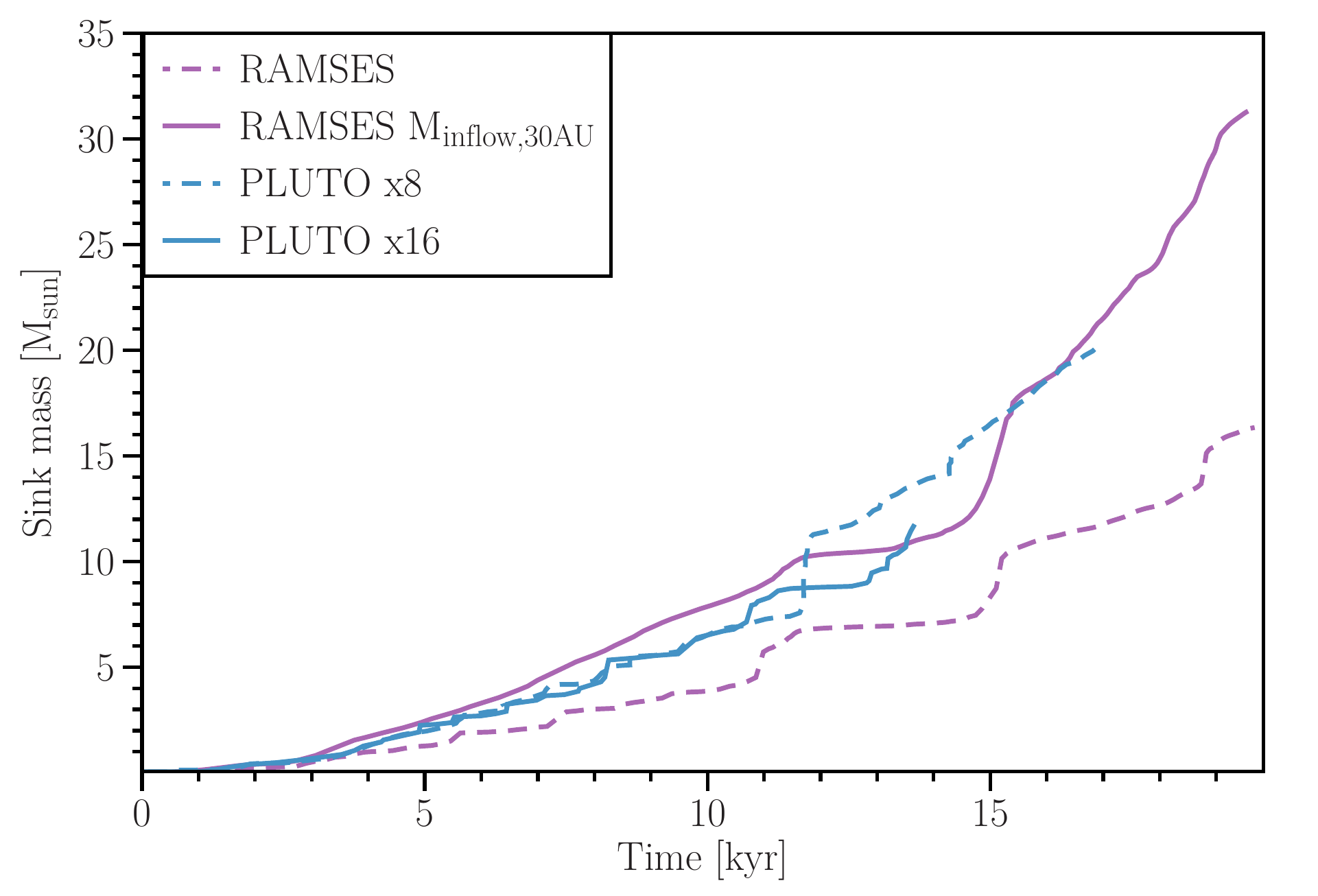} 
    \includegraphics[width=9cm]{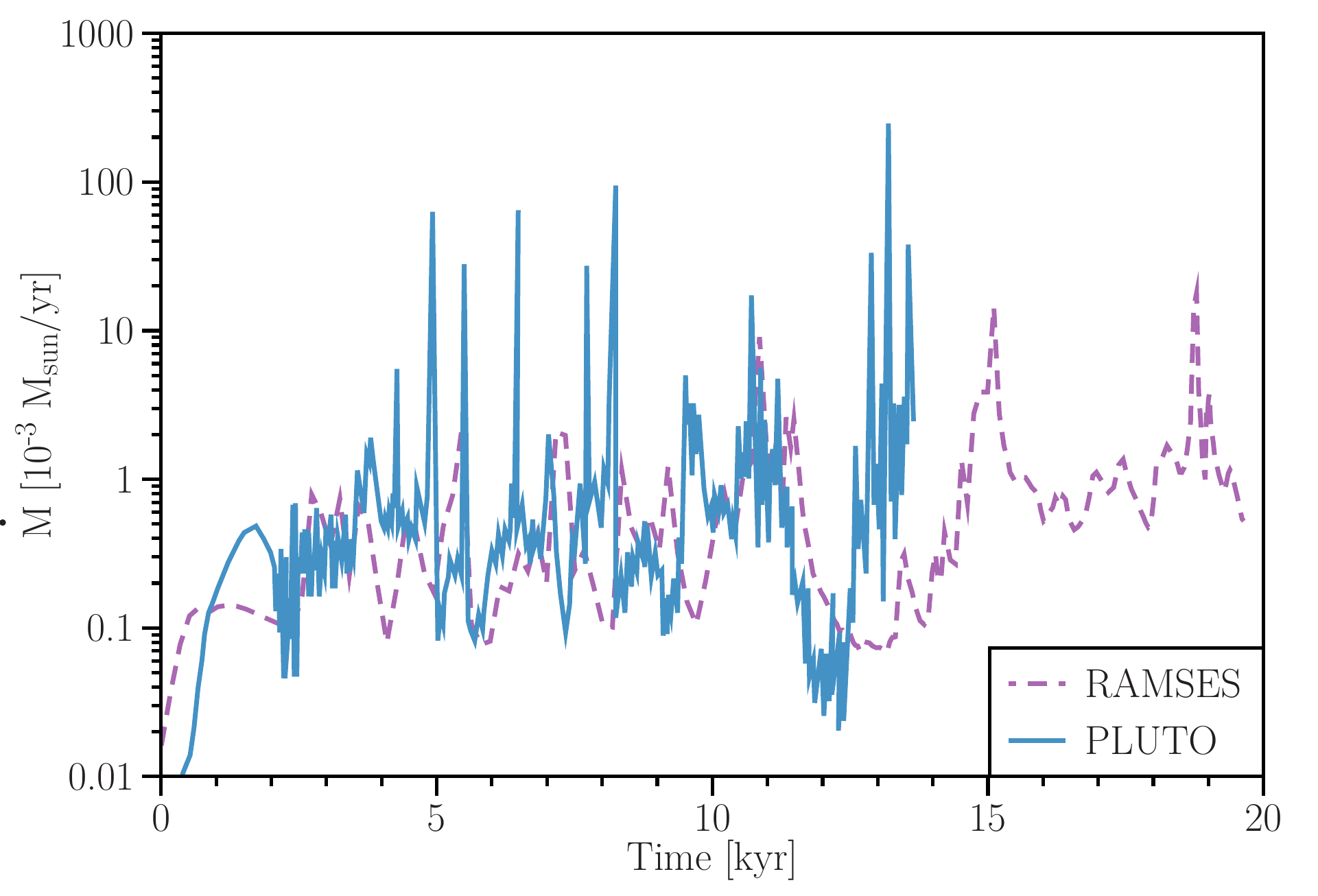}
    \caption{Left panel: sink mass as a function of time in the {\tt RAMSES} run against the {\tt PLUTOx8} and {\tt PLUTOx16} runs. The quantity $M_\mathrm{inflow,30AU}$ is the total mass flowing into a central sphere of radius $30$~AU, reproducing the accretion model of {\tt PLUTO} runs. Right panel: accretion rate onto the sink as a function of time in the {\tt RAMSES} run against the {\tt PLUTOx16} run.}
    \label{fig:tmsinkOK20}
\end{figure*}

First, let us note that the fragments formed following disk fragmentation in the \ramses{} run, at about $4$~kyr, are accreted eventually, similarly to the first fragments formed in {{\tt PLUTO} and shown in Fig.~\ref{fig:fragmentation}}.
{Nevertheless, some of their initial orbits are stable until fragment-fragment interactions promote accretion.}
Those fragments contribute to the mass growth of the central sink particle/cell, that we detail henceforth.

The sink mass in the {\tt RAMSES} run and in the {\tt PLUTOx8} and {\tt PLUTOx16} runs is plotted as a function of time in the left panel of Fig.~\ref{fig:tmsinkOK20}.
As in the rest of the paper, the instant $t=0$~kyr refers to the beginning of the simulation.
We find that the overall evolution of the sink mass is qualitatively similar, with several accretion bursts during which the sink gains more than one solar mass. This is due to the accretion of fragments, whose formation is observed in both studies (Sec.~\ref{sec:frag}). 
Meanwhile, the sink mass is always smaller in the {\tt RAMSES} run as compared to {\tt PLUTOx16} and {\tt PLUTOx8}, with a difference that can be as large as a factor of $2$. 
Possible explanations for this quantitative discrepancy are the density threshold for sink accretion or the ability for the gas in our simulation to leave the sink volume while it directly enters the sink mass in {\tt PLUTO} runs.
To check the former, we run a similar simulation but with a density threshold ${\approx}4$ times smaller, and the sink mass is nearly unchanged with $7.3\, \mathrm{M_\odot}$ at $t{\approx}13$~kyr instead of $7\, \mathrm{M_\odot}$.
This is consistent with \citealt{hennebelle_what_2020}, who report that the sink mass is marginally influenced by this threshold.
To check the latter, we integrate the total inflow mass into a sphere of radius $30$~AU centered on the sink.
Since we do not output every iteration of the run, the inflow rate used to compute the total inflow mass is smoothed using a median over seven outputs (the time step between outputs is about $0.1$~kyr).
This total inflow mass, labeled $M_\mathrm{inflow,30AU}$, is displayed in the left panel of Fig.~\ref{fig:tmsinkOK20}.
The estimate $M_\mathrm{inflow,30AU}$ compares well with {\tt PLUTOx8} and {\tt PLUTOx16}, except for the period between ${\approx}12$~kyr and ${\approx}15$~kyr, in which run {\tt PLUTOx8} exhibits an accretion burst.
Hence, the discrepancy between our sink mass and the {\tt PLUTO} runs likely comes from the difference in the accretion model: a density threshold in \ramses{} against a sink cell in {\tt PLUTO}.
In the former, gas is allowed to leave the sink volume without being accreted and the sink actually plays the role of boundary conditions for the disk \citep{hennebelle_what_2020}, while it is directly attributed to the sink mass in the latter and associated with boundary conditions on the hydrodynamical variables.

The accretion rate is displayed (in logarithmic scale) as a function of time in the right panel of Fig.~\ref{fig:tmsinkOK20}. It oscillates between a low accretion state with $\dot{M} {\sim}10^{-4} \mathrm{\, \mathrm{M_\odot} \, yr^{-1}}$ and a high accretion rate state with values higher than $\dot{M} {\sim}10^{-3} \mathrm{\, \mathrm{M_\odot} \, yr^{-1}}$. The accretion rate in the low state is similar to that found in {\tt PLUTOx16}. The high state gives a smaller accretion rate than what is reported by OK20. However, this state corresponds to the epochs when fragments are accreted, and fragments have typically the same mass in both studies (see Sec.~\ref{sec:frag}), so the quantitative difference partially comes from the different time bin (larger in our study) to compute the instantaneous accretion rate. Both studies exhibit a similar number of accretion bursts.

\subsection{Disk Keplerian motion}
\label{sec:ok20disk}

\begin{figure*}
\centering
    \includegraphics[width=9cm]{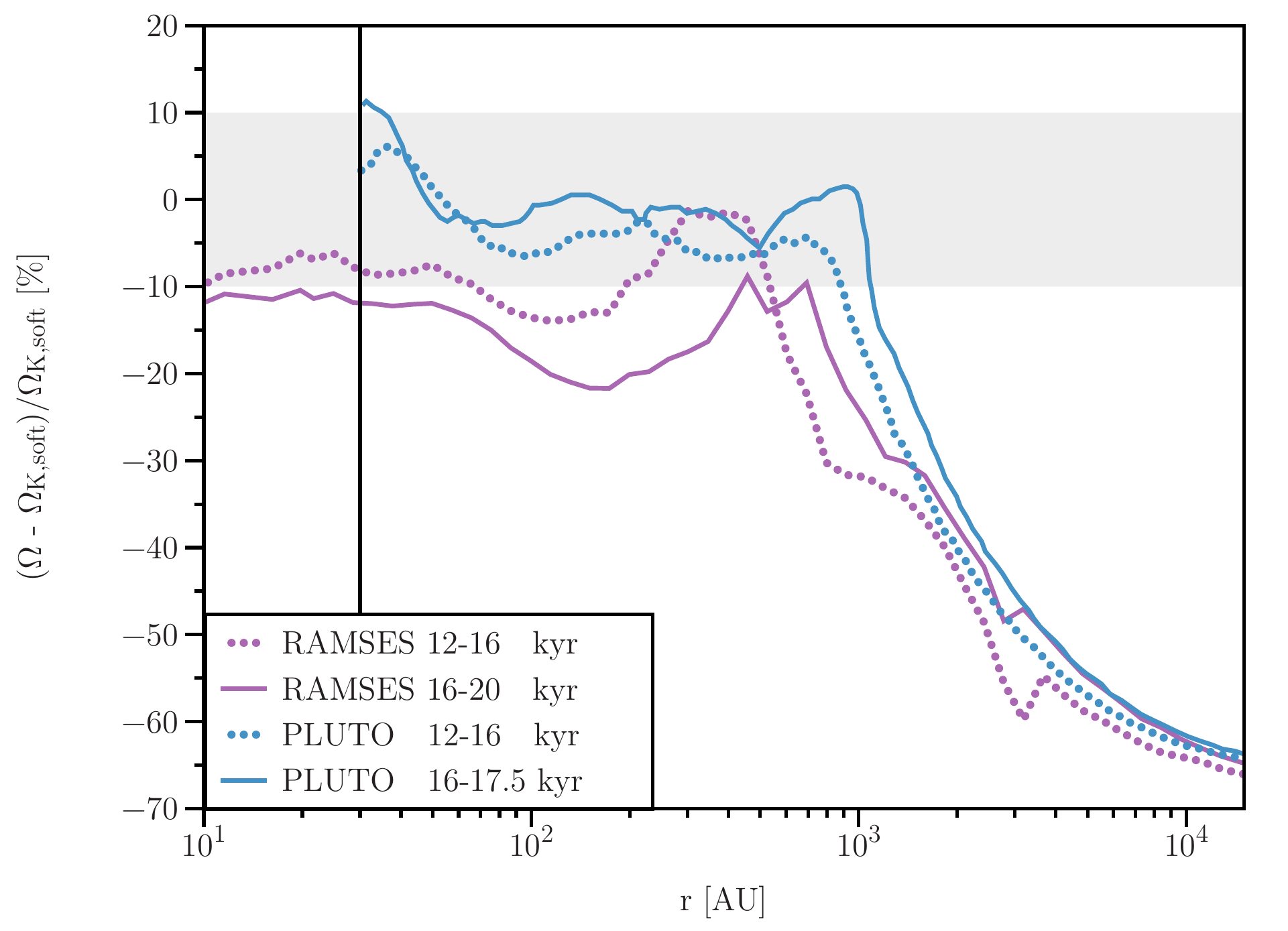} 
    \includegraphics[width=9cm]{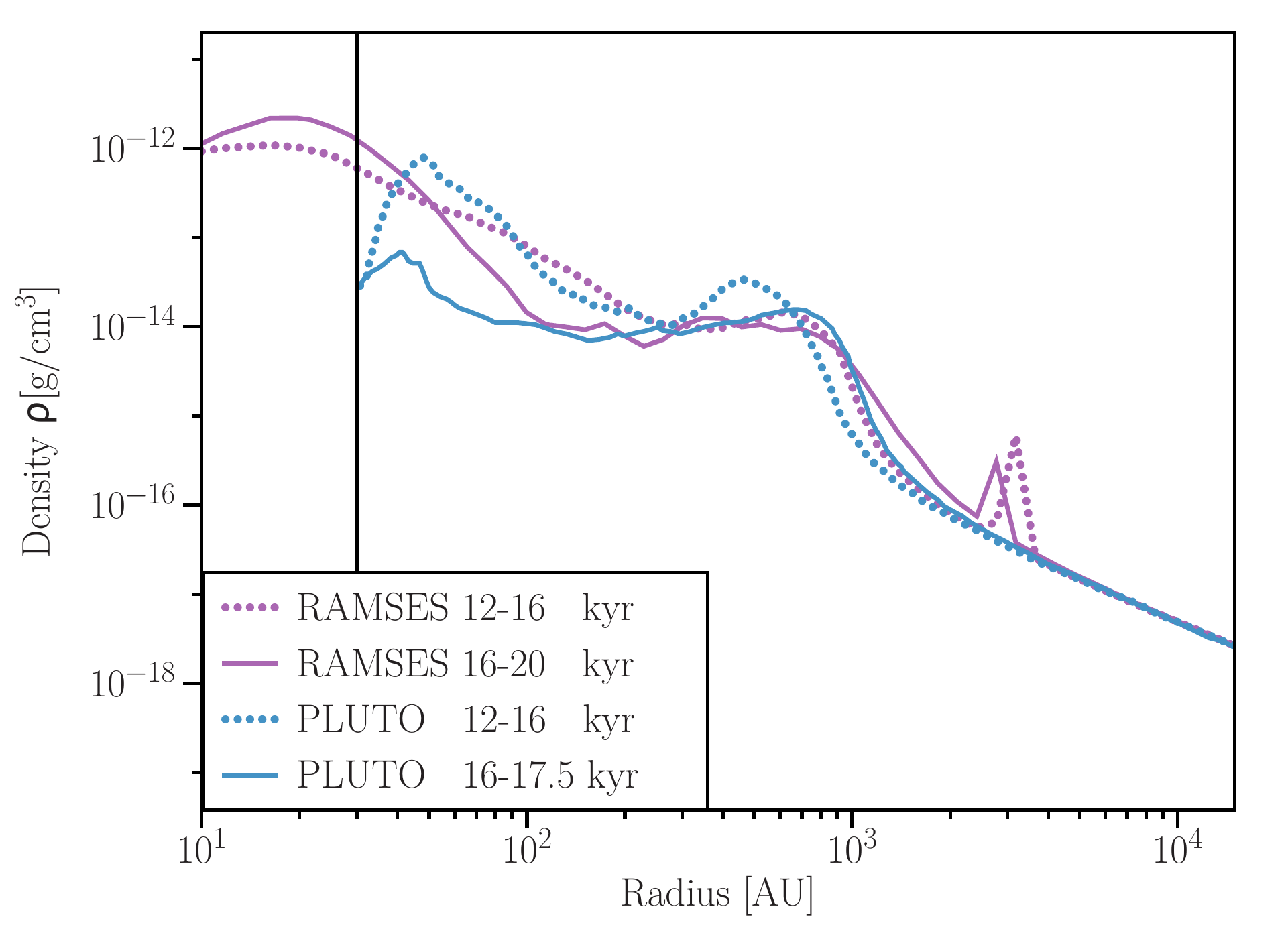}  
    \caption{Radial profiles of the deviation from Keplerian frequency (left panel) and the density (right panel) in the \ramses{} run and in {\tt PLUTOx8}. For each radius, we compute the azimuthal median and average it in time (as indicated in the plot legend).
    The vertical line indicates the sink cell radius in {\tt PLUTO} runs.}
    \label{fig:diskppties}
\end{figure*}

Figure~\ref{fig:diskppties} shows the radial profiles of the deviation from Keplerian frequency, defined using Eq.~\ref{eq:col_fragrkss} and the density in the {\tt RAMSES} and {\tt PLUTO} runs.
It can be seen in the left panel of Fig.~\ref{fig:diskppties} that the gas is slightly sub-Keplerian between $30$~AU and a few hundreds of AU, especially between $16$~kyr and $20$~kyr (down to ${\approx}-20\%$). On the opposite, the same region in {\tt PLUTOx8} shows a deviation between $-10\%$ and $10\%$. 
If the gas Keplerian motion should be used as a proxy for the disk radius, then the drop in deviation from Keplerian frequency points to a disk radius of ${\approx}550$~AU for the interval $[12,16]$~kyr and either ${\approx}70$~AU (using the first drop) or ${\approx}750$~AU (using the second drop) for the interval $[16,20]$~kyr, against ${\approx}900$~AU for the interval $[12,16]$~kyr and ${\approx}1000$~AU for the interval $[16,17.5]$~kyr in {\tt PLUTOx8}. 
Furthermore, we computed the thermal pressure gradient acceleration and we note that it is one order of magnitude too small to compensate this sub-Keplerian motion and ensure equilibrium.
Thus, the disk-like structure we obtain is not at equilibrium.
We observe that the sub-Keplerian region coincides with the region where the gas dynamics is dominated by interactions between fragments (collisions, gravitational interactions) and with the central star.
For instance, at ${\sim}18$~kyr, a clump (not hot enough to be detected as a fragment, see Sec.~\ref{sec:frag}) is partially disrupted in the vicinity of the central star.
Part of the debris stream is projected with radial velocities of the order of $10 \, \mathrm{km \, s^{-1}}$. The stream collides with the infalling, rotating gas, so the region swept by the stream only contains slowly-rotating, sub-Keplerian gas. 
Moreover, we find that the north-south symmetry has been broken, which we attribute to the multiple fragment collisions.

As shown in the right panel of Fig.~\ref{fig:diskppties}, the density in the central region is roughly in agreement with the findings of {\tt PLUTOx8}. We recall that we took the azimuthal median for each radius, hence the dense fragments and other non-axisymetries have been smoothed. 
This is true as long as more than $50\%$ of the cells within those bins are in a common state, referred to as the background disk state in OK20.
The drop in density around $1000$~AU, which could be used to define the primary disk as well, is found at a similar radius in both studies after $16$~kyr and with a slightly smaller radius in {\tt PLUTOx8} for $[12,16]$~kyr.
A first drop in density is also present at ${\approx}70$~AU for the $[16,20]$~kyr interval, at the same position as the drop in the deviation from Keplerian frequency previously reported.
This indicates the low density region produced by tidal disruptions of fragments.

As a side note, the spikes visible in the deviation from Keplerian frequency and in the density profiles at about $3000$~AU correspond to a fragment that has been ejected by fragment-fragment interactions around $7$~kyr. 
It appears to be gently falling back onto the central region.

\subsection{Impact of numerical methods on the Keplerian motion}

\subsubsection{Grid (de-)refinement}

In the following, we address the possible impact of numerical methods on the sub-Keplerian disk profile. To understand whether the AMR refinement and de-refinement could artificially prevent the disk from relaxing to quiescence, we run a similar simulation from the start but with a partially-fixed grid.
We use a geometrical criterion to fix the spatial resolution to $2.5$~AU up to ${\approx}200$~AU from the central star and to $5$~AU up to $400$~AU, in the disk plane and within a disk thickness of $30$~AU.
This results in a number of cells of size $2.5$~AU in a cylinder of radius $200$~AU and height $30$~AU centered onto the sink multiplied by ${\sim}4$ (from about 60000 cells to 240000 cells).
Further away, between radii of $200$~AU and $400$~AU, the number of cells of size $5$~AU is multiplied by ${\sim}6$ (from about 15000 cells to 90000 cells).
Additional refinement based on the standard Jeans length criterion is allowed. 
We obtain similar results in terms of Keplerian motion as in run {\tt RAMSES}. Hence, the sub-Keplerian motion is not due to the AMR grid.

\subsubsection{Axisymmetric gravitational potential on a Cartesian grid}
We also check whether this could come from a bad sampling of the density on a Cartesian grid, which should, in the case of an accretion disk, take a nearly axisymmetric distribution. 
{First, we obtain a nearly Keplerian disk until $t{\approx}12$~kyr and therefore the loss of Keplerian motion is unlikely to be caused by the grid being Cartesian.
For safety, we can check how is sampled a spherically-symmetric potential by computing} the gravitational potential of the pre-stellar core at $t=0$ (the bad sampling would be linked to the Cartesian grid and therefore should be already visible at $t=0$), and compare it to the analytical, textbook, value. We obtain an error of $2\%$.
{This suggests that the sub-Keplerian frequency, which reaches $-20\%$ is not a consequence of a bad sampling of the density distribution by the Cartesian grid.} This is consistent with the study of \cite{lyra_global_2008}, who ran simulations of disks in a Cartesian grid and were able to reproduce standard features (such as equilibrium) obtained in cylindrical and spherical codes (see also the code comparison by \citealt{de_val-borro_comparative_2006}).
Hence, the origin of the discrepancy between {\tt RAMSES} and {\tt PLUTO} is not attributable to the grid.

\subsubsection{Importance of the fragments dynamics and sink mass}
\label{sec:fragdyn}

Finally, to assess the dynamical origin of both the sub-Keplerian motion and the north-south asymmetry, we perform an identical simulation with a finest resolution of $10$~AU ({see Appendix~\ref{app:cvg}}).
Indeed, this resolution should slightly under-resolve the dense structures, and therefore reduce the impact of collisions. Moreover, it shifts the central sink accretion radius from $10$~AU to $40$~AU, so there is less gravitational energy available to fuel the tidal disruptions.
In this run (plot not shown here for conciseness), we find the vertical structure to remain roughly symmetric up to $20$~kyr. Furthermore, the rotation profile is closer to Keplerian rotation, with a smallest value of $-15\%$ to $-20\%$ {(see Fig.~\ref{fig:keplcvg2})}.
{Meanwhile, the sink mass evolution is similar, at late times, to that presented in Fig.~\ref{fig:tmsinkOK20} for run {\tt RAMSES} (see Fig.~\ref{fig:t_msink_cvg})}.
Hence, we conclude that the sub-Keplerian motion and north-south asymmetry are partially linked to the dynamics, i.e. collisions and tidal disruptions, in the disk-like structure.

Moreover, the total mass of the fragments during the interval $[16,20]$~kyr is around $6\, \mathrm{M_\odot}$ on average in \ramses{} (more details in Sec.~\ref{sec:frag}), while the central star mass is between $11\, \mathrm{M_\odot}$ and $16\, \mathrm{M_\odot}$. In run {\tt PLUTOx8}, the star mass is $17\, \mathrm{M_\odot}$ at $t=16$~kyr and the maximal mass of a fragment is between $3\, \mathrm{M_\odot}$ and $5\, \mathrm{M_\odot}$. Here, the central star dominates only marginally and locally, the total gravitational potential. Beyond a few hundred AU, the disk self-gravity dominates, thus it is more prone to gravitational instabilities (e.g. \citealt{kratter_gravitational_2016}) and less likely to reach quiescence.
Hence, the accretion model impacts the sink mass, as shown in Sec.~\ref{sec:msinkok20}, and it is certainly responsible for the discrepancy regarding the disk equilibrium (that is, Keplerian motion; indeed, thermal support is much smaller than rotation support).

\subsubsection{Discussion on the Keplerian motion}

Overall, we find that the \ramses{} run exhibits a more dynamical, or chaotical, disk-like structure than in OK20, where the disk is Keplerian and therefore at equilibrium between centrifugal and gravitational accelerations.
{No quiescence state is reached by the end of the simulated time, ${\approx}20$~kyr, unlike \pluto{} runs.}
This questions whether the gas orbiting the sink in the \ramses{} run should be labeled a (sub-Keplerian) disk and whether it would relax to a Keplerian disk state at later times, once the central mass becomes sufficiently massive.
A possibility to explain the discrepancy between OK20 and the present result lies in the mass growth of the central star and the fragments, because the disk stability increases with the star-to-disk mass ratio \citep{kratter_gravitational_2016}.
Indeed, we found a smaller stellar mass and slightly more massive fragments than OK20, resulting in a less stable disk than in their study.
The accretion model, namely a sink cell with inflow boundary condition in OK20, and a density threshold in the sink volume in our case, is likely responsible for this discrepancy in the sink mass.
In any case, it suggests a larger impact from the accretion model and the modeling of the central $30$~AU in radius than from grid effects (AMR and Cartesian).
Nevertheless, differences in the propagation of spiral waves, which contribute to angular momentum redistribution and subsequent accretion, on a spherical grid and on a Cartesian grid could also be at work to explain part of the aforementioned discrepancies.
Running the same simulations with a SPH code would allow for a complementary point of view. We leave such comparison studies to future work.

\subsection{Fragments tracking}

In the following, we study the temporal evolution of the fragments. First, we implemented the procedure presented in OK20 in order to detect fragments in {\tt RAMSES} outputs and identify them from one output to the next one.
In a nutshell, for a given output $n$ we extract the temperature map in the $(x=0,y,z)-$plane and we convolve it by a Gaussian filter to smooth the non-axisymmetries induced by spiral arms.
Then, we compute the azimuthal median profile of the temperature and, in each cell, retrieve the corresponding value to identify hot spots (i.e. zones of higher temperature than their corresponding azimuthal median).
A temperature threshold of $400$~K is then used to select the position of the remaining hot spots. The size of hot spots is set to $40$~AU in radius, as in OK20.
{With our finest resolution of $2.5$~AU and a refinement criterion based on the Jeans length, this ensures that the diameter of a fragment is sampled by $16$ cells.}
Once the position is obtained, we collect the data (central temperature, mass, density, velocity vector) of each fragment.
We use the output $n+1$ we extract the positions of new hot spots and compute the expected position of the old hot spots using a linear expansion in time, that is $r_\mathrm{exp} = r_\mathrm{n} + v_{r,n} (t_{n+1}-t_n)$ and $\phi_\mathrm{exp} = \phi_\mathrm{n} + v_{\phi,n}/r_n (t_{n+1}-t_n)$ for the radius and azimuthal angle $\phi$, respectively, where the subscript "exp" stands for "expected" and $n$ for the output number.
Comparing the positions of new hot spots with the surroundings of the expected position of old hot spots, we determine whether or not they correspond to the same physical fragment.
Finally, we manually checked the continuity of the orbits. 

\begin{figure}
\centering
    \includegraphics[width=9cm]{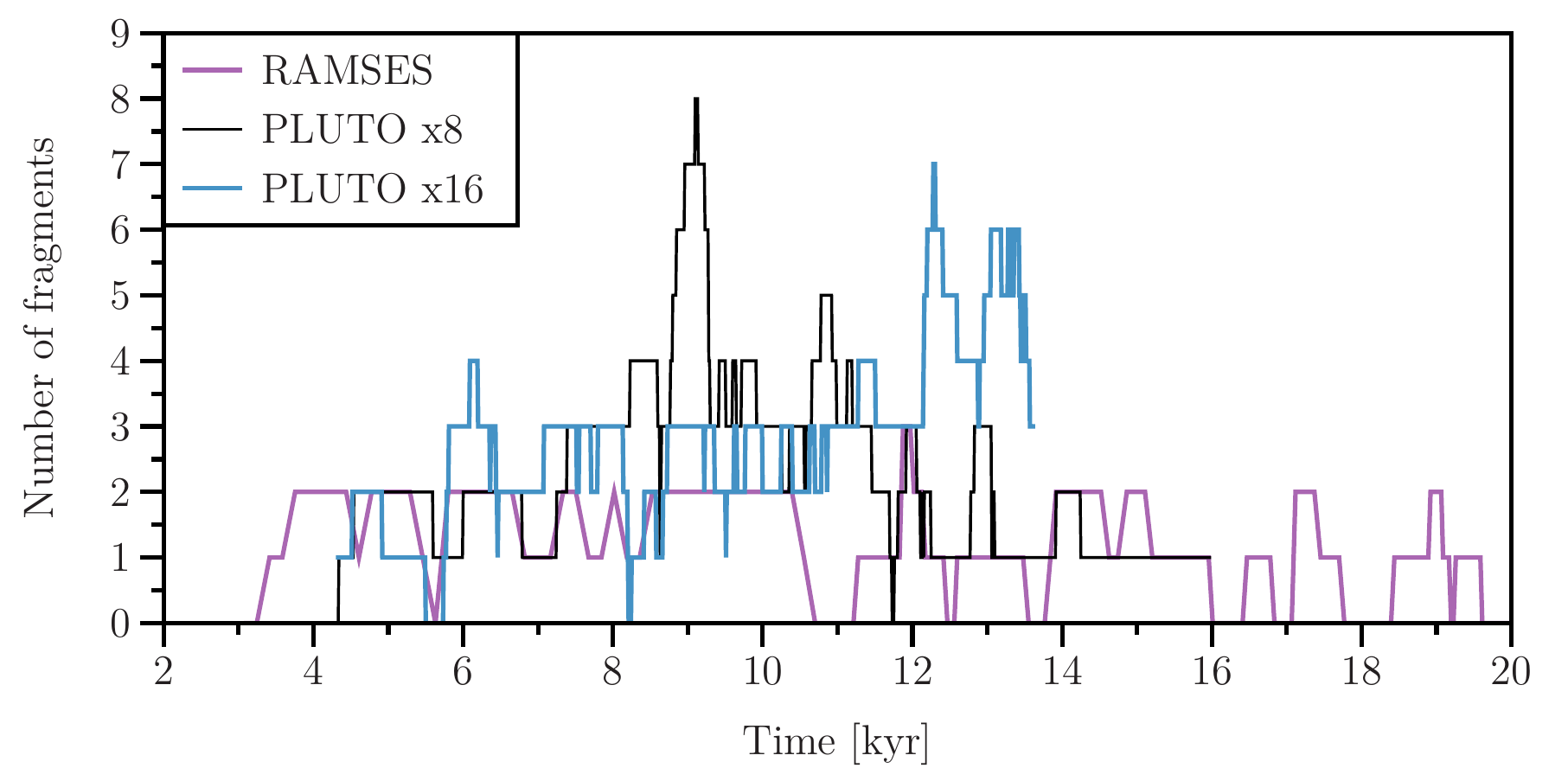}
    \caption{Number of fragments as a function of time. 
    Only fragments with a lifetime longer than $200$~yr are shown.}
    \label{fig:Nfrag}
\end{figure}

\subsection{Fragments properties}
\label{sec:frag}

{Fragments form within spiral arms or following spiral arm collisions, in \ramses{} as in \pluto{}, as already reported in several studies (see e.g. \citealt{bonnell_new_1994}, \citealt{bonnell_massive_1994-1}).
They} orbit around the central star on eccentric orbits and eventually get destroyed by various processes: tidal disruption after approaching the central star (e.g. fragments $\#1$ and $\#8$), thermal expansion (e.g. fragment $\#13$) or shear when transported over a spiral arm (e.g. fragment $\#14$).
{These processes occur in both {\tt RAMSES} and {\tt PLUTO} runs.}

{First of all, the number of fragments is in correct agreement between \ramses{} (and among the \ramses{} runs, see the convergence study in Appendix~\ref{app:cvg}) and \pluto{}.
Figure~\ref{fig:Nfrag} shows the number of fragments detected as a function of time.
The number of fragments is of the same order of magnitude and varies between $0$ and $2$ in \ramses{} and $0$ and $4$ in \pluto{}.
The few times when the number of fragments is very distinct in \ramses{} and \pluto{} are associated with a transient peak of fragment formation/destruction, e.g. at ${\approx}9$~kyr, run {\tt PLUTOx8}. 
Such a peak (as also visible at ${\approx}12$~kyr in {\tt PLUTOx16}) shows how non-linear the formation and destruction of fragments are, as new collisions between fragments and spiral arms can occur and either trigger new fragment formation or lead to their destruction/mergers, reducing their number.
Hence, rather than focusing our study on the exact fragment number at a given time, we are interested in statistical trends and more importantly, on what is the physical origin of those trends.
We note that the total fragment mass (see below) does not follow the peak behaviour reported above, suggesting that is it associated with low-mass fragments whose feedback on the background disk properties is small, hence we treat this event as transient and not decisive for the rest of the simulation.
Interestingly, a fragmentless disk, product of simultaneous fragment-destruction events, is reported in {\tt PLUTOx8} at $11-12$~kyr and in \ramses{} at $10-11$~kyr.
A plausible outcome of such a fragmentless period would be for the primary disk to enter a quiescent phase as a result of the reduced activity in the disk, and provided that the central star is massive enough to stabilize it. This is not the case in \ramses{} but is the case at late times in {\tt PLUTOx8}, when the temperature increase due to stellar irradiation in the innermost parts of the disk also contributes to the stabilization.
Overall, except for transient events of fragment formation/destruction, \ramses{} and \pluto{} yield very consistent results with respect to the number of fragments present on the disk as a function of time.
}

\begin{figure}
\centering
    \includegraphics[width=9cm]{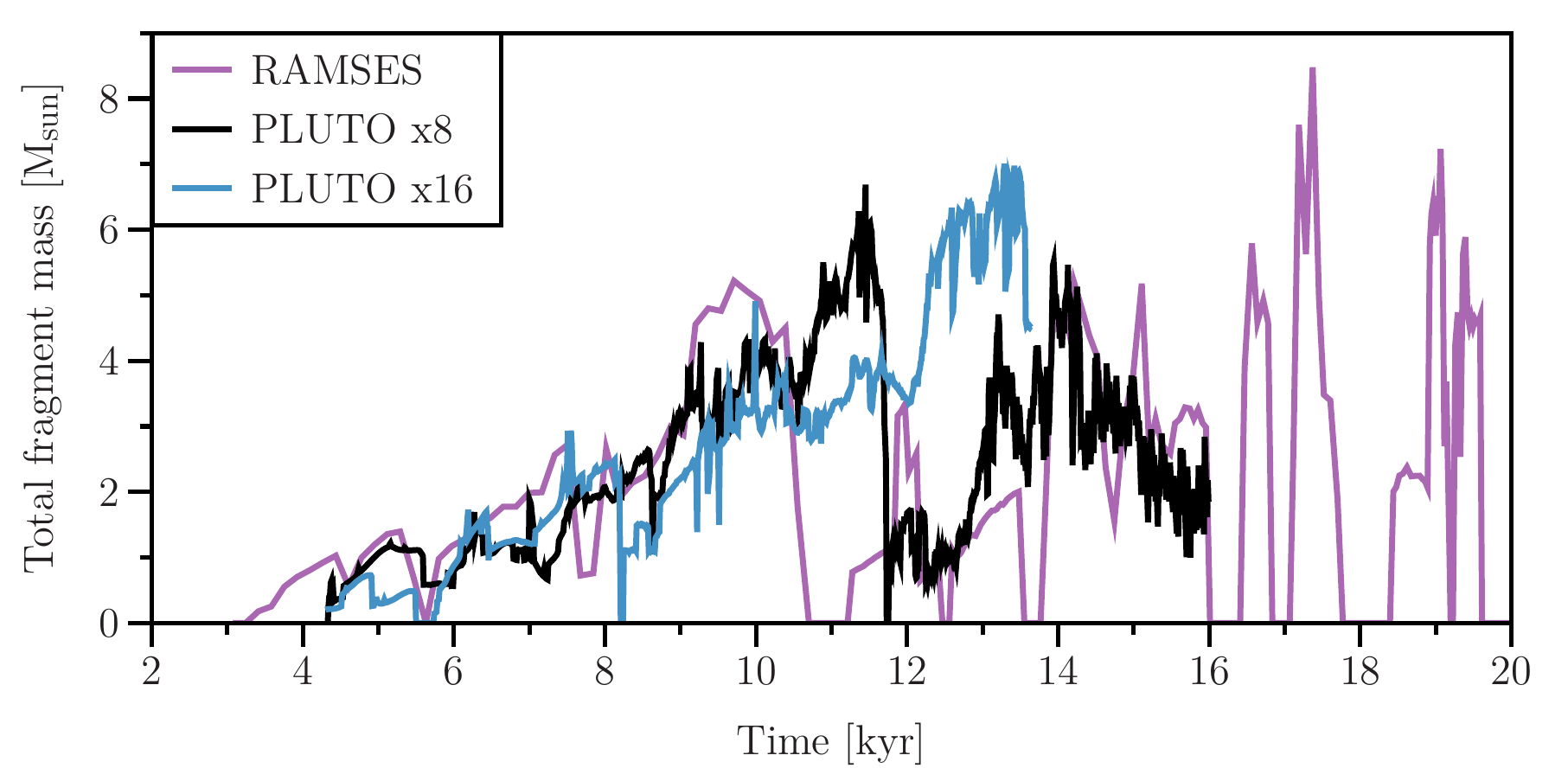} 
    \includegraphics[width=9cm]{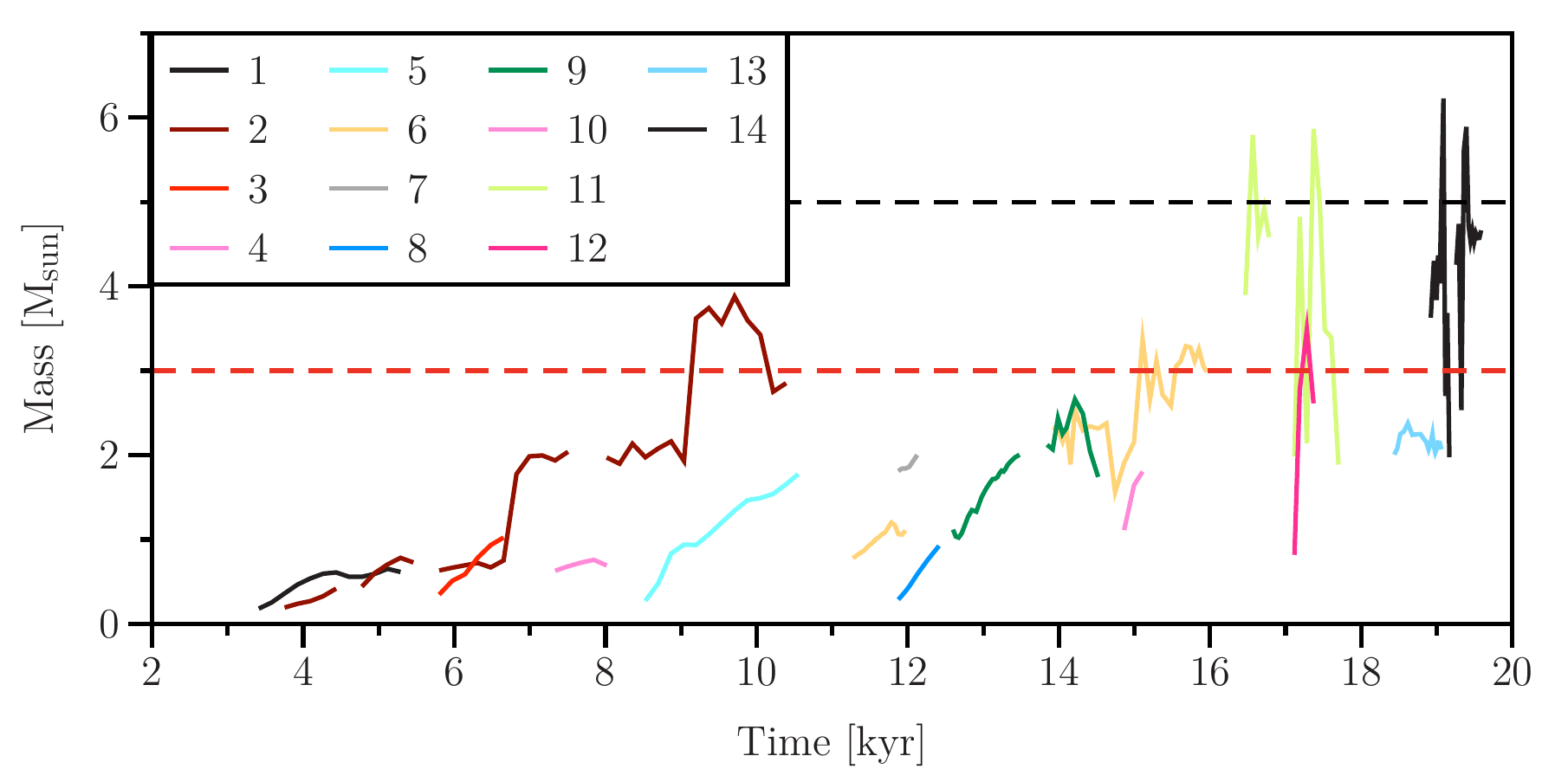}  
    \caption{Total mass in all runs (top) and individual mass in \ramses{} (bottom) of the fragments as a function of time. Only fragments with a lifetime longer than $200$~yr are shown. 
    {For visibility and comparison, in the bottom panel, the black and red dashed lines show the highest fragment mass in run {\tt PLUTOx8} and {\tt PLUTOx16}, respectively.}}
    \label{fig:Mfrag}
\end{figure}

{We now turn to the mass of fragments.}
Figure~\ref{fig:Mfrag} shows the {total (top panel) and invididual (bottom panel)} mass of the fragments as a function of time. 
{The total fragment mass smoothly increases with time and is, on average, ${\sim}2-2.5 \, \mathrm{M_\odot}$ in both codes.
Up to ${\sim}10$~kyr, the total fragment mass evolution is very similar in all runs.
This is understandable as the mass budget for fragments is linked to the growth of the primary disk.
After $10$~kyr, the total fragment mass abruptly decreases in \ramses{} (except for a finest resolution of $10$~AU, see Appendix~\ref{sec:app_fragcvg}) and {\tt PLUTOx8} as an event of simultaneous fragment destruction occurs, as previously reported.
}
{Individually, the fragment masses} range from a fraction of a solar mass up to six solar masses. In comparison, the most massive fragment is $5\, \mathrm{M_\odot}$ in run {\tt PLUTOx8} and $3\, \mathrm{M_\odot}$ in run {\tt PLUTOx16}, {as indicated by the dashed lines.}
The high mass reached by fragments $\#11$ and $\#14$, between $3\, \mathrm{M_\odot}$ and $6\, \mathrm{M_\odot}$, suggests that they had the potential to form rapidly a massive companion.
Moreover, we notice a trend for forming more massive fragments at later times than early times, in agreement with OK20.
Indeed, the initial mass enclosed within a radius $r$ increases with the radius as $r^{1.5}$ so there is more gas available then.
Another possibility would be the build-up of the accretion structure around the primary sink, but as discussed above (Sec.~\ref{sec:ok20disk}), such a structure is not at equilibrium in the \ramses{} run, unlike that of OK20.

\begin{figure}
\centering
    \includegraphics[width=9cm]{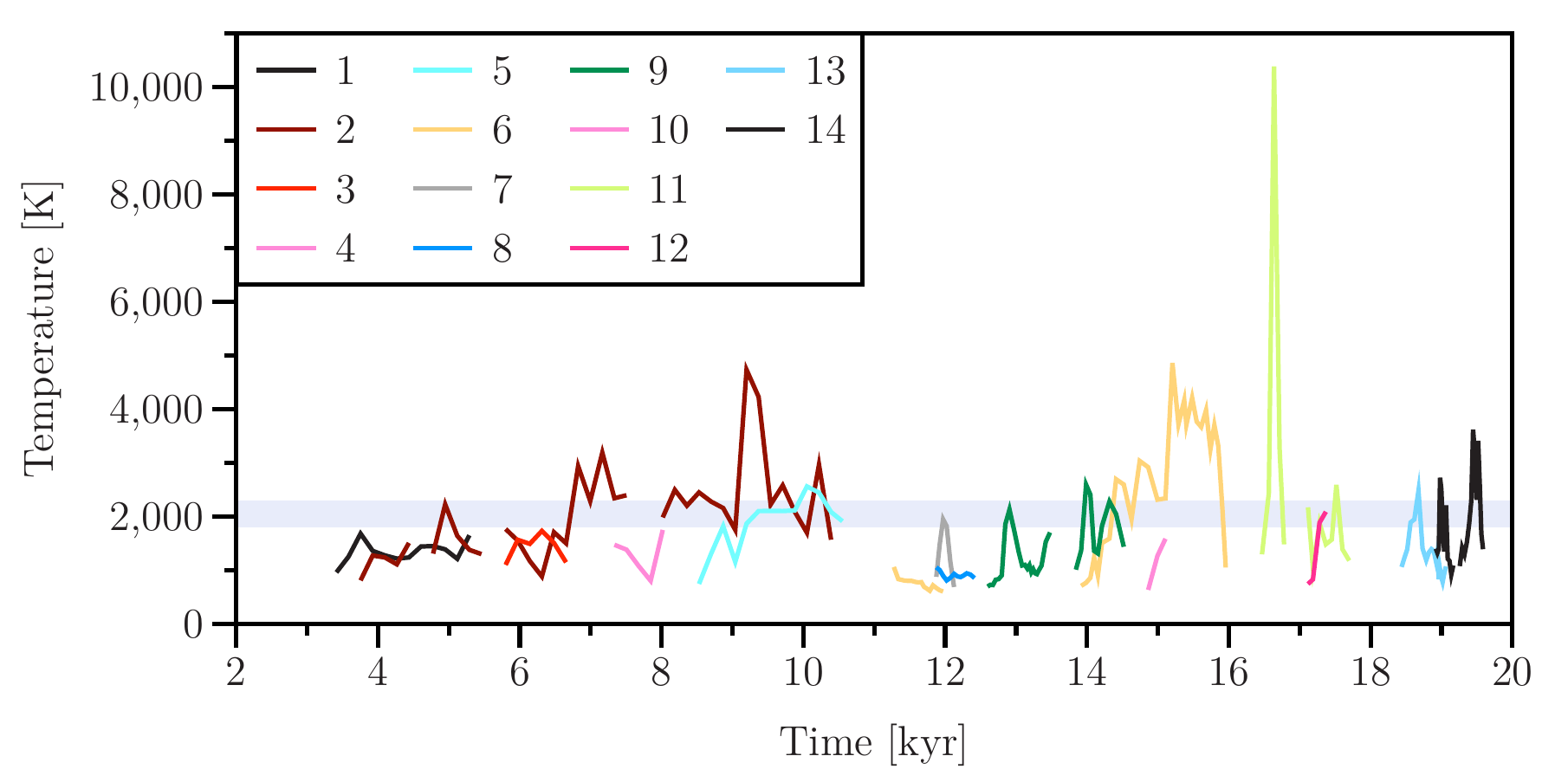}  
    \caption{Central temperature of the fragments as a function of time. Only fragments with a lifetime longer than $200$~yr are shown.}
    \label{fig:Tfrag}
\end{figure}

{Let us now focus on the fragment temperature.}
Figure~\ref{fig:Tfrag} shows the temperature of the fragments as a function of time. The gray band indicates the $\mathrm{H_2}$ dissociation limit $T\sim 2000$~K, as in OK20. Fragments reaching this limit are expected to undergo second collapse and form second Larson cores (see e.g. \citealt{vaytet_simulations_2013} and \citealt{bhandare_birth_2020} for a dedicated study on the second core formation).
We report that $9$ fragments reach this limit, against $10$ in run {\tt PLUTOx16} and $4$ in run {\tt PLUTOx8}.
Except for fragment $\#11$, whose temperature is due to a collision event that compresses the gas adiabatically - because it is optically thick - up to $\rho{\approx}5\times 10^{-9} \mathrm{g\, cm^{-3}}$, the fragments temperature lies in a range consistent with {\tt PLUTO} runs.
The fragment temperature appears correlated with the fragment density (see Fig.~\ref{fig:Mfrag}, the radius being fixed) suggesting adiabatic heating for all fragments. 

\begin{figure}
\centering
    \includegraphics[width=9cm]{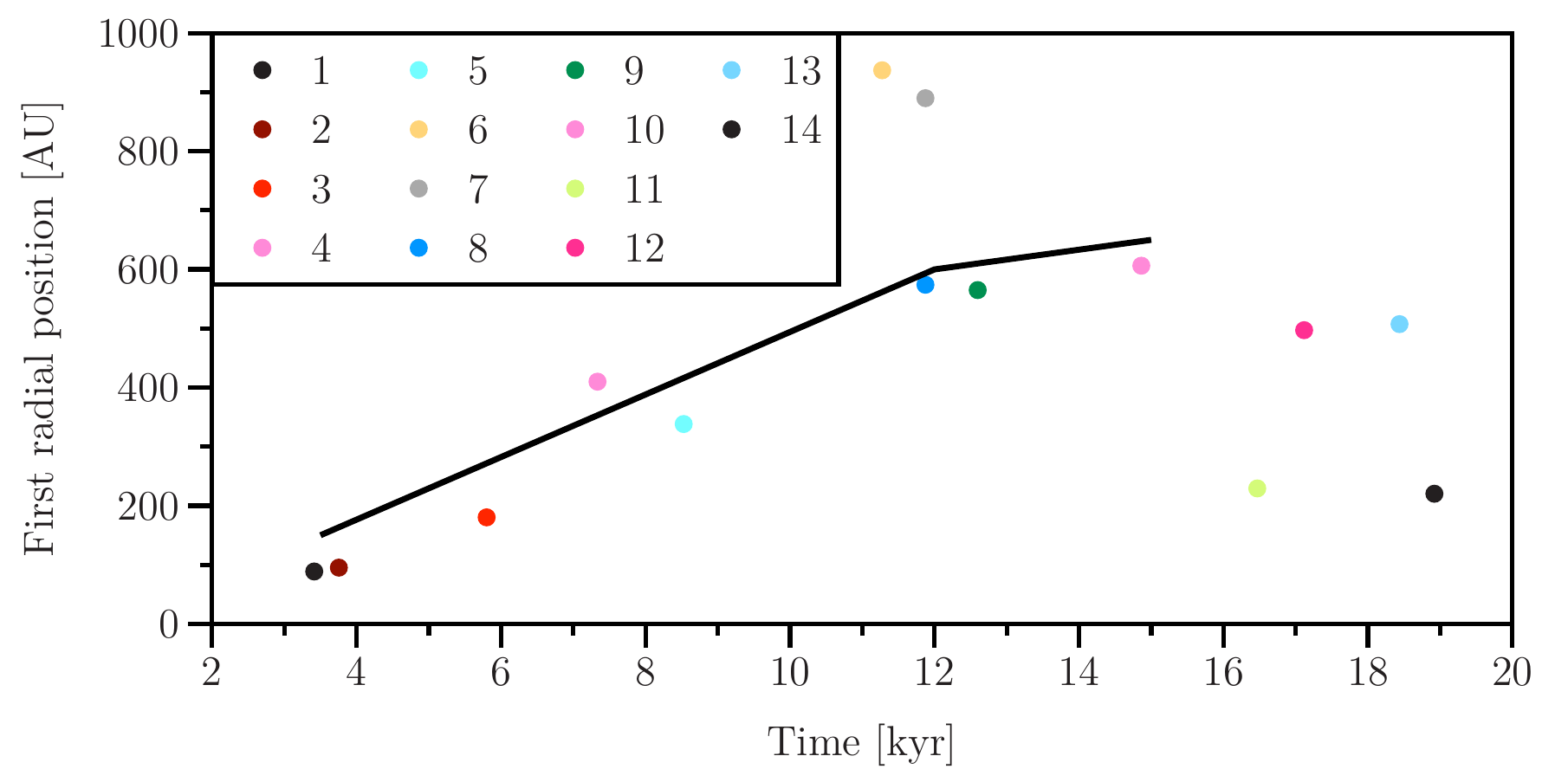}
    \caption{Distance of newly born fragments to the central star, as a function of time. Only fragments with a lifetime longer than $200$~yr are shown.
    {The black curve indicates the maximal distance of newly born fragments in the {\tt PLUTOx8} run.}}
    \label{fig:radfrag}
\end{figure}

Figure~\ref{fig:radfrag} shows the distance to the primary sink when fragments are detected, as a function of time. As the disk grows, fragments can form at larger distances from the star. Nevertheless, the formation of fragments at smaller radii is not suppressed (see e.g. fragments $\#11$ and $\#14$). Fragments $\#6$ and $\#7$ form from the collision of two flows and migrate outwards while reaching the temperature threshold for detection, which explains the large distance at which they form.
{Except for those two fragments,  the distance is consistent with the maximal distance of newly born fragments in the {\tt PLUTOx8} run, plotted as the black curve.}

Overall, the fragments properties are in agreement between the \ramses{} and the \pluto{} runs.
{This indicates that, despite distinct radiation-hydrodynamical approaches, both codes reach a satisfying agreement at modeling the local thermodynamical behaviour of gaseous fragments in a massive protostellar disk.}

\section{Conclusions}
\label{sec:ccl}

{We presented the self-gravity-radiation-hydrodynamical simulations of the collapse of a massive pre-stellar core performed with the Cartesian AMR code \ramses{}, that we compare to the highest resolutions runs of \cite{oliva_modeling_2020}, performed with a modified version of the  code \pluto{} using a grid in spherical coordinates.}

{As a preliminary step, we chose the \ramses{} numerical setup for comparison to \pluto{}.
We compared two \ramses{} runs, one with a unique, central, fixed sink particle, and the other without any initial sink but the possibility to form sinks later-on.
Those two runs lead to qualitatively distinct systems: the former leads to a centrally-condensed system, the latter to a multiple stellar system born out of Toomre instability seeded by the Cartesian grid.
As the divergence is inherited from the first fragmentation phase, it shows how crucial is fragmentation in the innermost regions of the cloud for the future evolution of the system, and in this problem, the numerical caveats introduced by the use of sink particles and by the grid. 
It is not clear yet which of the two is the most realistic one.
For future studies, the issue of the instability seeded by the grid could be overcome by introducing additional, dominant, perturbations (see e.g. {\citealt{boss_fragmentation_1979},}  \citealt{commercon_protostellar_2008}) or by accounting for the inflow from larger scales (e.g. \citealt{vazquez-semadeni_hierarchical_2016}, \citealt{padoan_origin_2020}) while still resolving disk scales \citep{lebreuilly_protoplanetary_2021}.
{Additionally, turbulence in the massive pre-stellar core could be included to match the observational constraints of some pre-stellar cores (e.g. \citealt{beuther_formation_2007}, \citealt{bontemps_fragmentation_2010}, \citealt{palau_early_2013}, \citealt{girart_dr_2013}, \citealt{fontani_magnetically_2016}, \citealt{nony_detection_2018}) and may introduce density and velocity perturbations dominating over numerical ones.}}

{To perform the code comparison in the context of a centrally-condensed system, we chose the \ramses{} run with a unique, central, initial sink particle, as it compares qualitatively with the runs presented in \cite{oliva_modeling_2020}.
In the early phases of the collapse, gas free-falls towards the central region.
The central density increases and switches from isothermality to adiabaticity.
Additional infall of rotating gas triggers the formation a rotationally-supported disk whose border is the location of an accretion shock.
A good agreement between \ramses{} and \pluto{} is reached regarding the timeline of these events, as well as on the core and disk radius.
A "rotational bounce" occurs in the \ramses{} run and forms a density bump, while the \pluto{} runs show the formation of axisymmetric shocks on the same timescales and propagating through the disk.
This discrepancy might be due to the fine treatment of self-gravity, pressure gradient and centrifugal acceleration while conserving linear and angular momentum in a tiny (${<}50$~AU in radius) portion of the ($20000$~AU in radius) cloud.
In  \ramses{} the disk fragments at the location of the bump, which is Toomre-unstable, and in \pluto{} the early disk evolves two spiral arms which fragment due to their high density (low Toomre-parameter value).
These events occur on similar timescales: this is the first fragmentation era.
The accretion disk progressively grows around the central star.
It is consistent with Keplerian rotation in both codes, from its formation epoch to $12$~kyr.
Using the Keplerian frequency as a criterion to define the disk size, both codes show a good agreement: a few percent difference.}

{The accretion disk grows with time and the star gains mass, while fragments form continuously in the disk.
We detected and followed those fragments forming around the central star via their temperature.
The number of fragments reaching the $\mathrm{H_2}$ dissociation limit and their overall temperature is in agreement between the two codes, as well as their formation position.
Some of them are slightly more massive in the {\tt RAMSES} runs than the fragments formed in the {\tt PLUTO} runs ($6$ against $5 \, \mathrm{M_\odot}$ for the most massive fragments formed in each code), but the two codes find an overall satisfying agreement on the fragment properties.
This indicates that, in the present radiation-hydrodynamical frame with self-gravity, the local thermodynamics of fragments is consistent between \ramses{} and \pluto{}.}

{In the disk dynamical epoch, covering its growth and fragmentation, the disk is found to be sub-Keplerian over hundreds of AU in \ramses{}, while it is Keplerian in \pluto{}.
We tested several hypotheses to explain this result: the outcome of numerical methods (grid refinement and de-refinement, bad sampling of the nearly axisymmetric gravitational potential on a Cartesian grid) and the relevance of fragments dynamics and of the sink mass.
We found that the disk sub-Keplerian motion originates from tidal disruption of fragments and collisions, which strongly modify the velocity field in the disk region.
It produces spiral arms sweeping off the gas, slowing down the infall and reducing the amount of rotating gas around the central star.
Furthermore, the dynamics of fragments has more impact in a system where the disk-(and fragments)-to-star mass ratio is high.
In fact, this ratio is higher in \ramses{} than in \pluto{}.
Indeed, while fragments are slighlty more massive in \ramses{}, the stellar mass is about twice smaller as compared to \pluto{}.
We hypothesize that this discrepancy originates from the stellar accretion model. 
Indeed, in \ramses{}, the sink only accretes gas above a given, user-defined (in the simulations presented here), density threshold. 
There is an additional constraint of not accreting more than $10\%$ of the amount of gas above this threshold at each time step.
Meanwhile, the accretion procedure in \pluto{} represents a $100\%$ efficiency with no density threshold: the gas entering the sink cell is accreted.
When mimicking the accretion model of \pluto{} (namely, all the gas entering the sink volume is accreted) in the \ramses{} outputs, we reproduce quite successfully the accretion history of the \pluto{} runs.
Apart from this discrepancy, we mention nevertheless that the accretion history is qualitatively similar in \ramses{} and \pluto{} and consists both of continuous accretion and accretion bursts associated with fragments being accreted.
The order of magnitude of the stellar mass and of the accretion rate is similar in both codes.
However, as we show, a factor of two on the stellar mass is crucial for the dynamics of the massive protostellar disk at such early stages of the protostellar evolution phase.}
{This suggests that the detail of accretion mechanisms, based on star-disk interaction, are not only important for the stellar growth but also for the disk equilibrium and for the properties of the subsequent multiple stellar system.}

{We conclude that the differences in the initial fragmentation phase, potentially triggered by numerical choices (the grid, the use of sink particles), have more impact on the final multiplicity of the system than the choice of the code itself, between \ramses{} and \pluto{}{, when smooth initial conditions are employed}.}

\begin{acknowledgements}
      RMR thanks the referee for helping improving this manuscript. 
      RMR thanks Peggy Varniere and Ugo Lebreuilly for fruitful discussions. RMR acknowledges Patrick Hennebelle for his insights on numerical angular momentum conservation. This work was supported by the CNRS "Programme National de Physique Stellaire" (\emph{PNPS}). 
      This work has received funding from the French Agence Nationale de la Recherche (ANR) through the project COSMHIC (ANR-20-CE31-0009).
      The numerical simulations we have presented in this paper were produced on the \emph{CEA} machine \emph{Alfv\'en} and using HPC resources from GENCI-CINES (Grant A0080407247). The visualisation of \ramses{} data has been done with the \href{https://github.com/nvaytet/osyris}{OSYRIS} python package, for which RMR warmly thanks Neil Vaytet.
      G.A.O.-M. acknowledges financial support by the Deutscher Akademischer Austauschdienst (DAAD), under the program Research Grants - Doctoral Programmes in Germany, and financial support from the University of Costa Rica for the obtention of his doctoral degree.
      RK acknowledges financial support via the Emmy Noether and Heisenberg Research Grants funded by the German Research Foundation (DFG) under grant no.~KU 2849/3 and 2849/9.
\end{acknowledgements}

\bibliographystyle{aa} 
\bibliography{Zotero} 

\newpage
\begin{appendix}

\section{On the density bump/ring formation}
\label{app:bump}

\subsection{{Dependence on numerical and physical parameters}}

In the following we perform several checks to assess the robustness of the density bump/ring structure reported in the main text with respect to thermodynamics, physical and numerical parameters.
As shown in Sec.~\ref{sec:setup}, removing the sink particle in order to deal with self-gravity-hydrodynamics only gives a ring instead of a density bump.
The ring interior is made of low-density, sub-Keplerian material, but the processes of density bump and ring formation are similar.
Hence, we further focus on how the ring forms because it removes any influence from the central sink.

In the absence of a sink particle, to check whether the ring formation is a purely dynamical effect or if it is linked to the thermodynamics, we turn-off the FLD module and switch to a barotropic equation of state, using the density threshold for adiabaticity as $10^{-13} \mathrm{g \, cm^{-3}}$, as in \cite{cha_formation_2003} : the outcome is unchanged. 
Finally, we switch to an isothermal equation of state: the rebound occurs on the most central and densest region because of the pressure gradient (as in \citealt{larson_collapse_1972}, \citealt{black_evolution_1976}). This time, the pressure gradient has been built only by the density and not by the temperature (as in the barotropic and FLD cases, where pressure increases along with the temperature).
This confirms the initial dynamical role of pressure gradient in forming this protostellar ring.
We mention that, due to the aforementioned importance of pressure gradient forces, the name "rotational bounce" has been challenged by \cite{narita_characteristics_1984}.

In addition, we have run other simulations to explore the role of numerical and physical parameters on ring formation.
We find this ring to be a robust feature with respect to the Riemann solver (Lax-Friedrich and HLLD, \citealt{miyoshi_multi-state_2005}, which is less diffusive), to the initial rotation profile in the inner $30$~AU (no rotation, solid-body rotation, and differential rotation), {to the initial density profile in the inner $30$~AU (plateau or power-law)} and to the numerical resolution (from $10$~AU to ${\simeq}0.3$~AU resolution), making the angular momentum diffusion origin less plausible.
Changes performed within the inner $30$~AU were motivated by this size corresponding to the sink cell in OK20, which add degrees of freedom in our simulations.
Finally, it is certainly dependent on (other) initial conditions, as we do not report it in \cite{mignon-risse_new_2020} nor \cite{mignon-risse_collapse_2021}.

\subsection{Code comparison}

A density bump was present in {\tt PLUTO} runs (Fig.~\ref{fig:r_rho_25kyr}), formed from accretion shocks, and rapidly fragmented into two pieces (Fig.~\ref{fig:fragmentation}).
Hence, the origin of the structure is likely different from the explanation above.
Indeed, rotation only plays a role in the bump formation in {\tt PLUTO} by flattening the density towards the midplane, the first hydrostatic core is initially located within the sink cell, the density (contributing to the pressure gradient) is affected by the inner, zero-gradient, boundary condition, and the gas cannot exit the sink cell to feed the structure, unlike our simulation.
In a Cartesian AMR simulation such as the {\tt RAMSES} run, the most central region suffers from low angular resolution of the orbital elements. 
Indeed, the region concerned with the ring is very small compared to the size of the system: its formation and evolution is certainly strongly affected by minor errors in angular momentum conservation (\citealt{larson_collapse_1972}, \citealt{tscharnuter_collapse_1975},
\citealt{tohline_ring_1980},
\citealt{gingold_fragmentation_1983}).
Future work is required here.

The ring formation we report is, however, reminiscent of early analytical works \citep{tohline_ring_1980}, numerical studies with SPH codes (e.g. \citealt{bonnell_formation_1994}, \citealt{cha_formation_2003}, \citealt{hennebelle_protostellar_2004}), 2D axisymmetric calculations \citep{narita_characteristics_1984}, and nested grids \citep{matsumoto_fragmentation_2003}; see also the review by \cite{larson_physics_2003} and the comparison study performed by \cite{bodenheimer_comparison_1979}.
Further studies on such ring formation are needed, as the ring (or density bump in presence of a central sink particle) is very unstable and can naturally lead to multiple system formation \citep{norman_fragmentation_1978}.
Any other angular momentum transport mechanism than those included in this study could also prevent ring formation in astrophysical systems.

\section{Convergence study in {\tt RAMSES}}
\label{app:cvg}


{We consider three runs with a finest AMR level of $13$ ({\tt Low-res} run), $14$ ({\tt Mid-res} run) and $15$ ({\tt High-res} run) to perform a convergence study.
The {\tt High-res} run corresponds to the fiducial {\tt RAMSES} run presented in the main text.
The aforementioned AMR levels result in physical finest resolutions of $10$~AU, $5$~AU and $2.5$~AU, respectively.}

\subsection{Disk rotation support and stellar mass growth}

\begin{figure}
\centering
    \includegraphics[width=9cm]{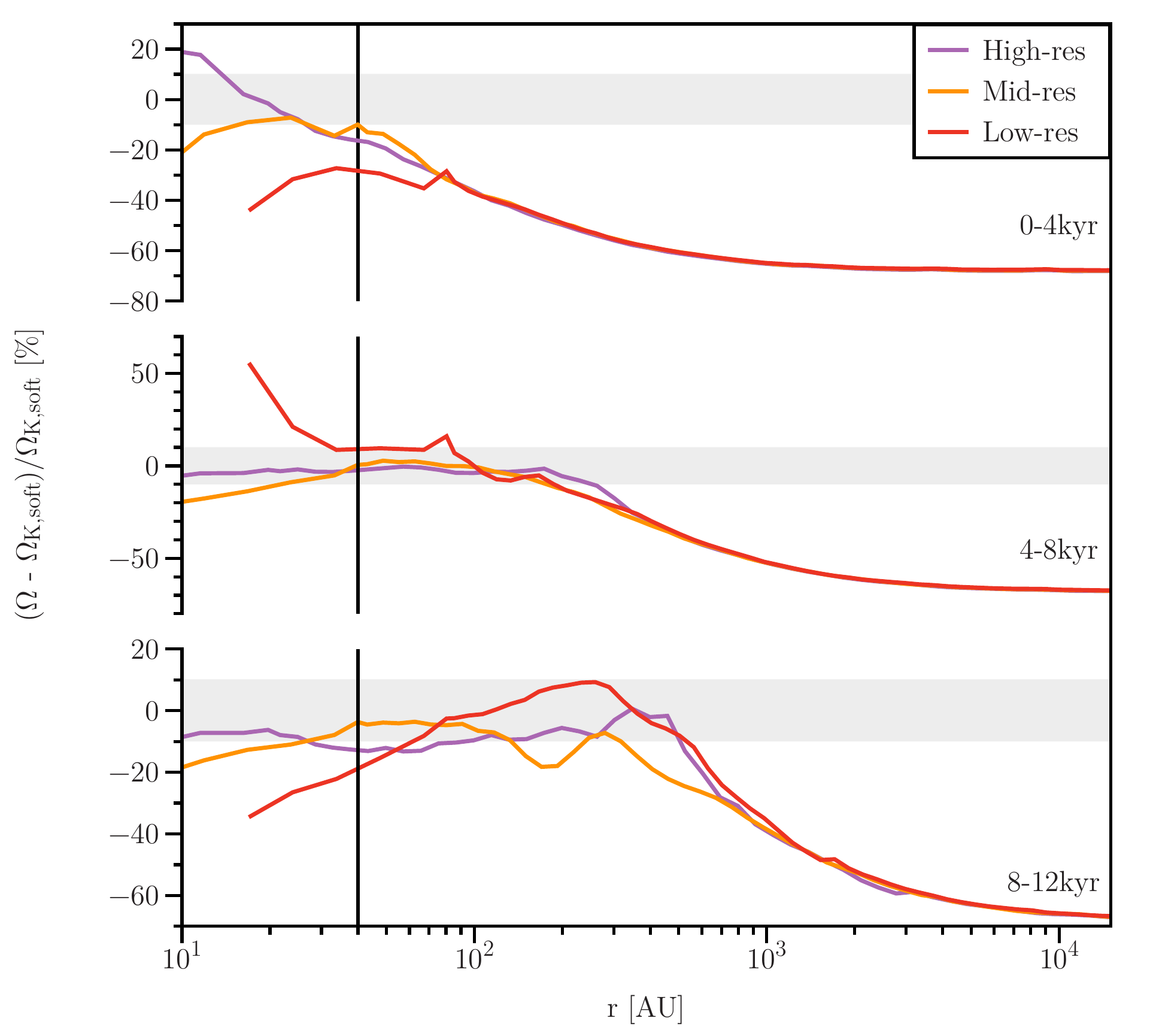} 
        \caption{Radial profiles of the deviation from Keplerian frequency for various resolutions in the disk formation epoch, covering, from top to bottom, the $[0,4]$~kyr, $[4,8]$~kyr, and $[8,12]$~kyr time intervals. The vertical line indicates the sink particle accretion radius in the AMR513 run.}
    \label{fig:keplcvg1}
\end{figure}

{The deviation from Keplerian frequency during the disk formation and growth phase is shown in Figure~\ref{fig:keplcvg1}.
At small radii, the same trend is visible in almost all runs and epochs with a decreasing deviation from Keplerian frequency as $r$ goes to zero (super-Keplerian frequency at small radius is always transient).
This effect is simply shifted to larger radii at lower resolution because the sink accretion radius is multiplied by two as the refinement is reduced by one level.
There is an overall agreement on the disk Keplerian motion. 
We note the presence of a sub-Keplerian region around $200$~AU in the {\tt Mid-res} run in the $8-12$~kyr epoch.}

\begin{figure}
\centering
    \includegraphics[width=9cm]{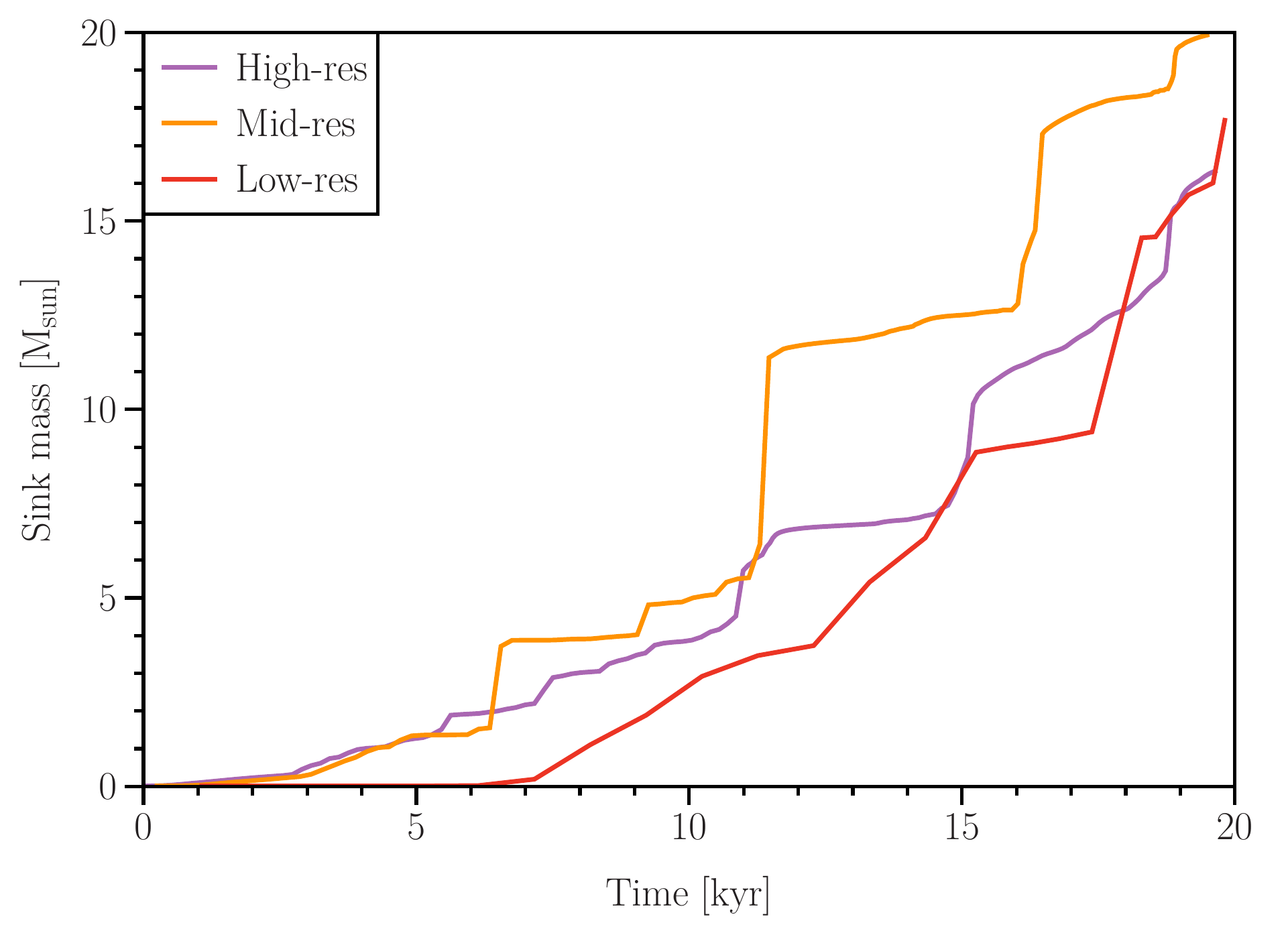} 
        \caption{Sink mass as a function of time for various resolutions.}
    \label{fig:t_msink_cvg}
\end{figure}

{The mass history of the central object is shown in Fig.~\ref{fig:t_msink_cvg}. 
The final stellar mass is consistent from one run to the other with a deviation of about $15\%$.
However, a comparison at a given time is made difficult by major accretion bursts, e.g. in the {\tt Mid-res} run at $11$~kyr.
This event is connected to the non-Keplerian region reported above in the $[8,12]$~kyr epoch, showing once again the importance of the dynamics of fragments for the computation of the Keplerian frequency.
Noticeably, the stellar mass growth in the {\tt Low-res} run is delayed compared to the other runs.
In fact, the first core density does not reach the accretion threshold before the ring starts forming and expelling gas from the center, hence starving the sink for a few kyr until the ring fragments.
However, the fragments originating from the ring are eventually accreted by the star so it finally catches up with the {\tt High-res} run accretion history.}

\begin{figure}
\centering
    \includegraphics[width=9cm]{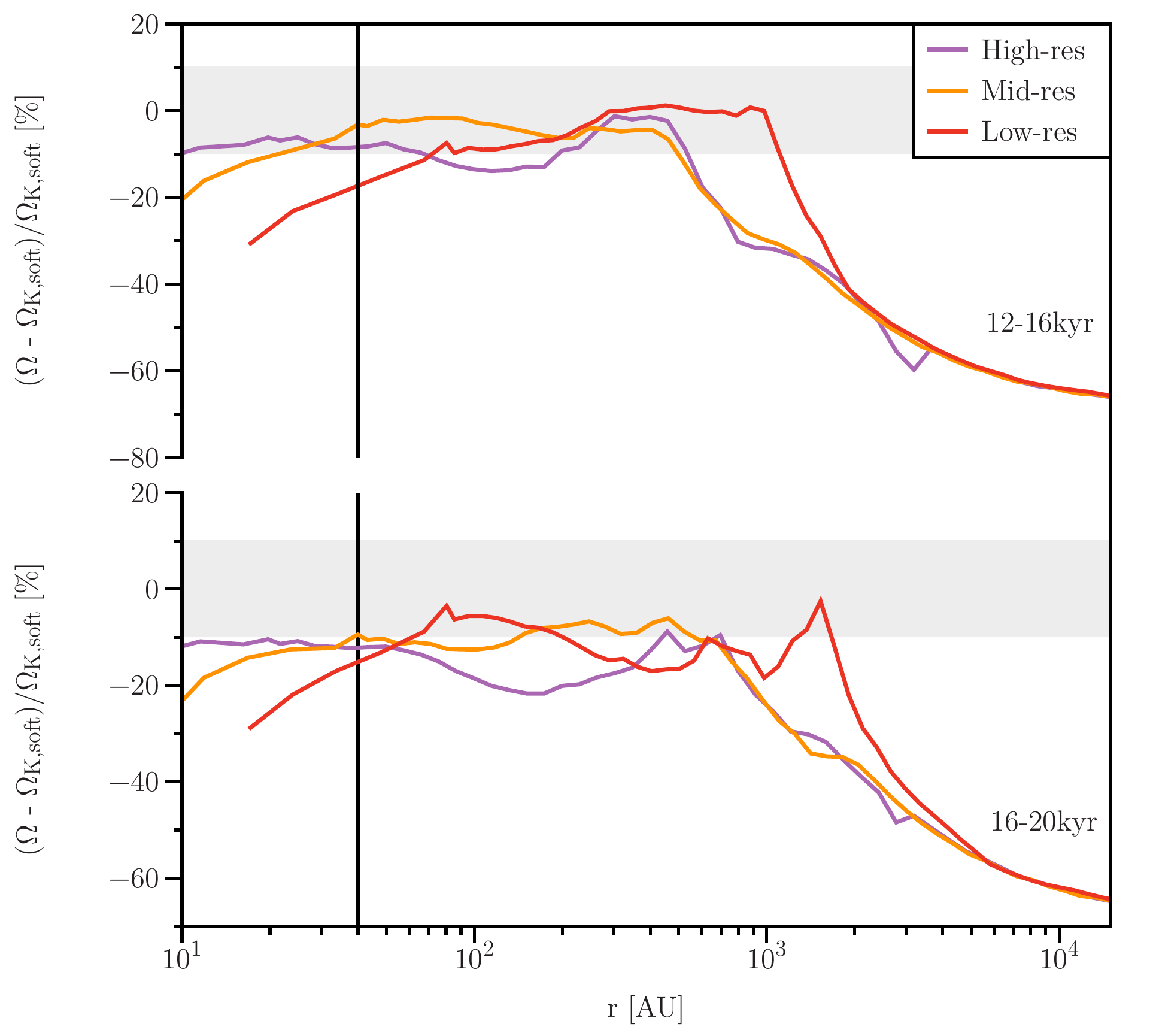} 
        \caption{Radial profiles of the deviation from Keplerian frequency for various resolutions in the disk dynamical epoch, covering, from top to bottom, the $[12,16]$~kyr and $[16,20]$~kyr time intervals. The vertical line indicates the sink particle accretion radius in the AMR513 run.}
    \label{fig:keplcvg2}
\end{figure}

{The deviation from Keplerian frequency during the disk dynamical epoch is shown in Figure~\ref{fig:keplcvg2}.
Qualitatively, runs {\tt Low-res} and {\tt High-res} exhibit at least once the presence of non-Keplerian regions in what would be expected to be the disk region.
This occurred before for run {\tt {\tt Mid-res}}, between $8$ and $12$~kyr, when a major accretion burst occurred while the sink mass was still rather small ($5\, \mathrm{M_\odot}$).
There is nevertheless a good agreement between the {\tt Mid-res} and {\tt High-res} runs for the disk radius in the $[12,20]$~kyr interval.
The disk radius is larger in the {\tt Low-res} run, while until the $[8,12]$~kyr epoch it was consistent with the {\tt Mid-res} and {\tt High-res} runs.
As already underlined in the main text, this is attributable to the dynamics of fragments.
At lower resolution, there is less gravitational energy to tap in during a close encounter with the central sink since it cannot get as close to the sink as in higher resolution runs.
Hence, there is less kinetic energy for the fragments to destabilize the disk Keplerian motion.
The regions that are non-Keplerian, from $400$ to $\gtrsim 1000$~AU in run {\tt Low-res}, indeed coincide with the region in which the two densest disk fragments are located, but their drops in Keplerian frequency are shallower than in the {\tt Mid-res} and {\tt High-res} runs.
Moreover, a fragments disruption does not just affect the Keplerian motion at a given time. When it occurs, the post-disruption region is not centrifugally-supported so the infalling gas from larger scale penetrates this region until it reaches the centrifugally-supported disk and contributes to its build-up there.
Applying this reasoning to the {\tt High-res} run, part of the infalling material at epoch $[16,20]$~kyr eventually contributes to bringing the $[100-400]$~AU region back to Keplerian motion instead of increasing the radius of the global disk-like structure.
This explains why the disk is larger in run {\tt Low-res}.}

\subsection{Fragments properties}
\label{sec:app_fragcvg}

\begin{figure}
\centering
    \includegraphics[width=9cm]{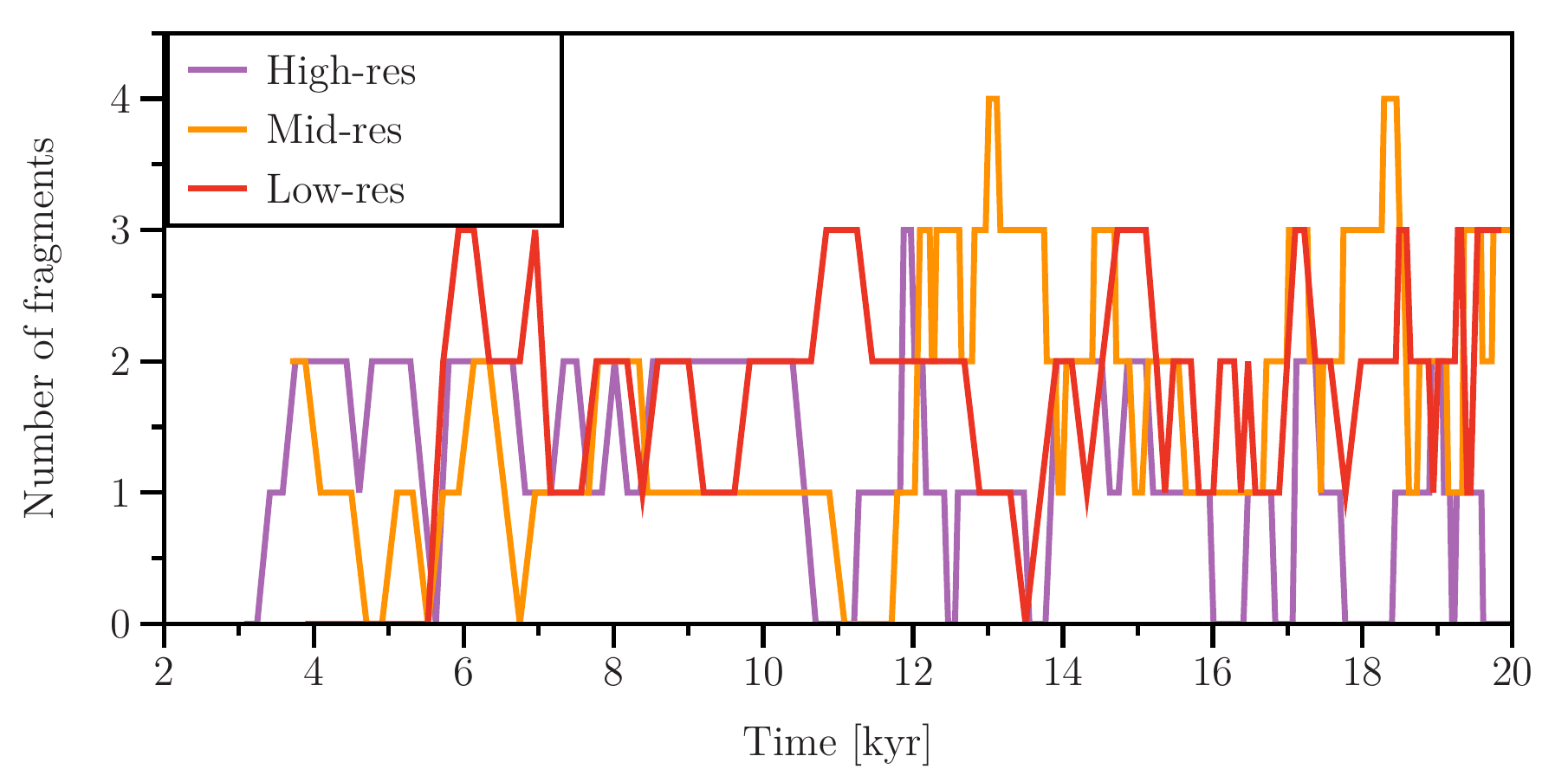} 
    \includegraphics[width=9cm]{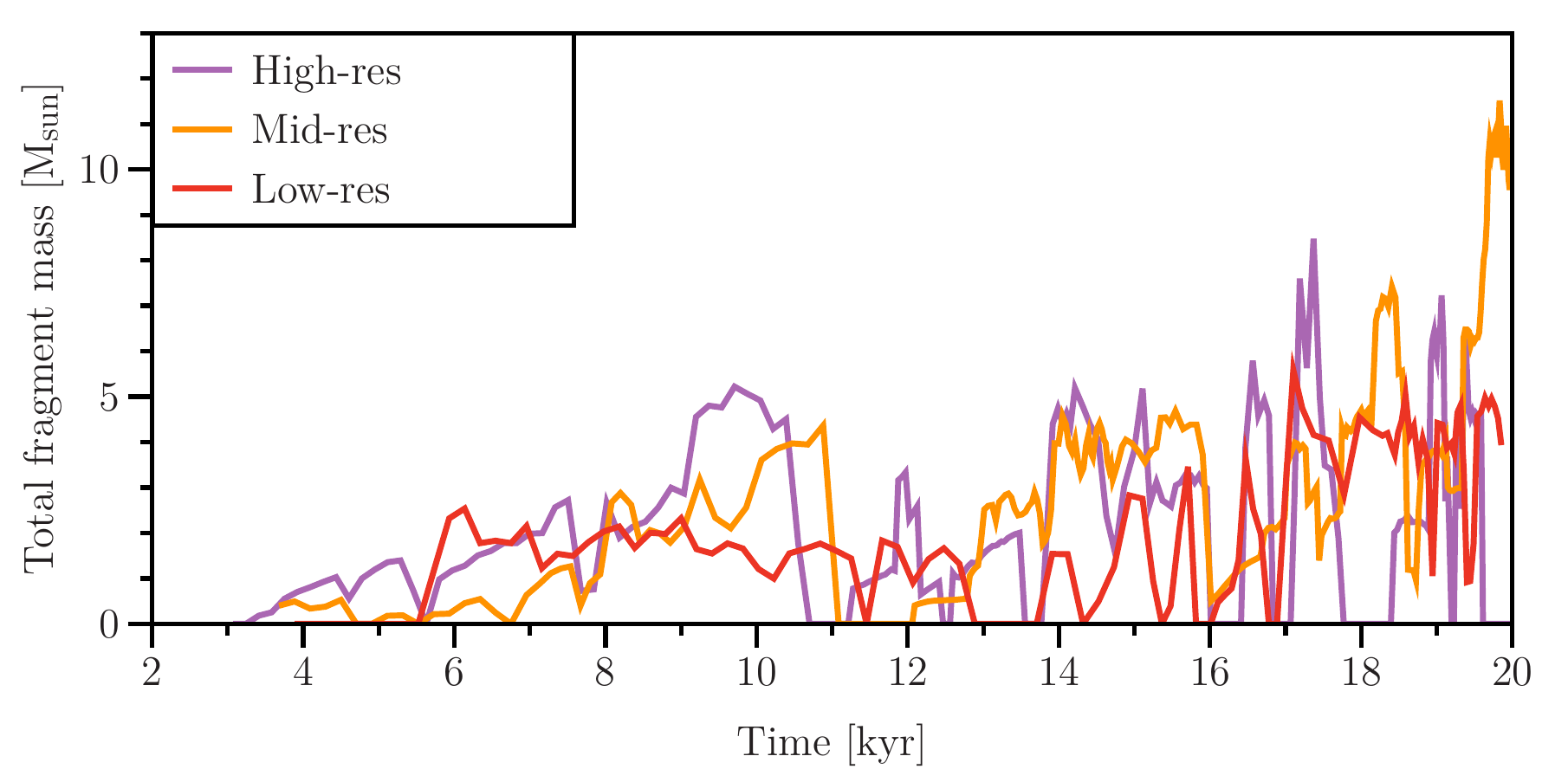} 
    \includegraphics[width=9cm]{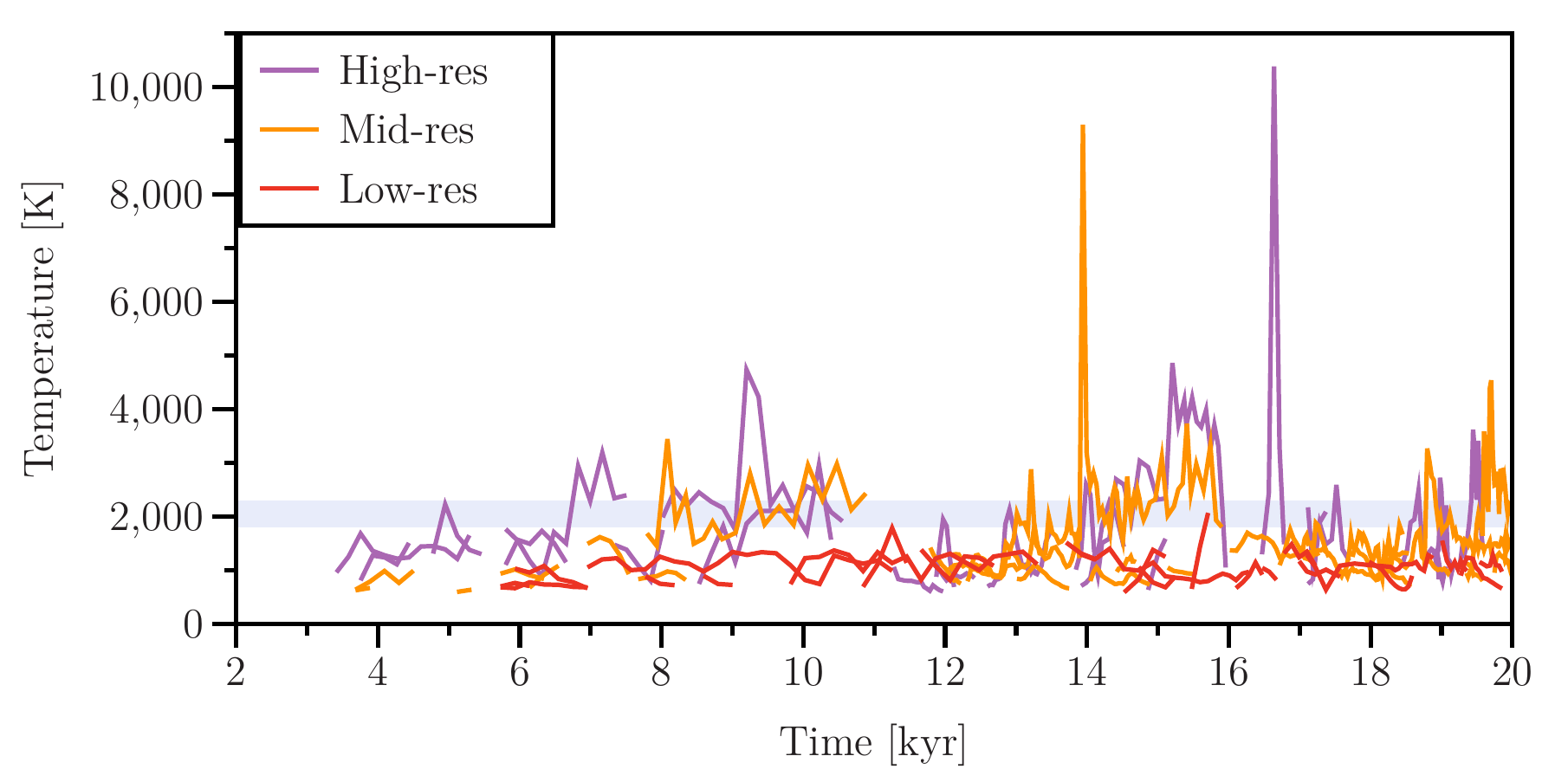} \includegraphics[width=9cm]{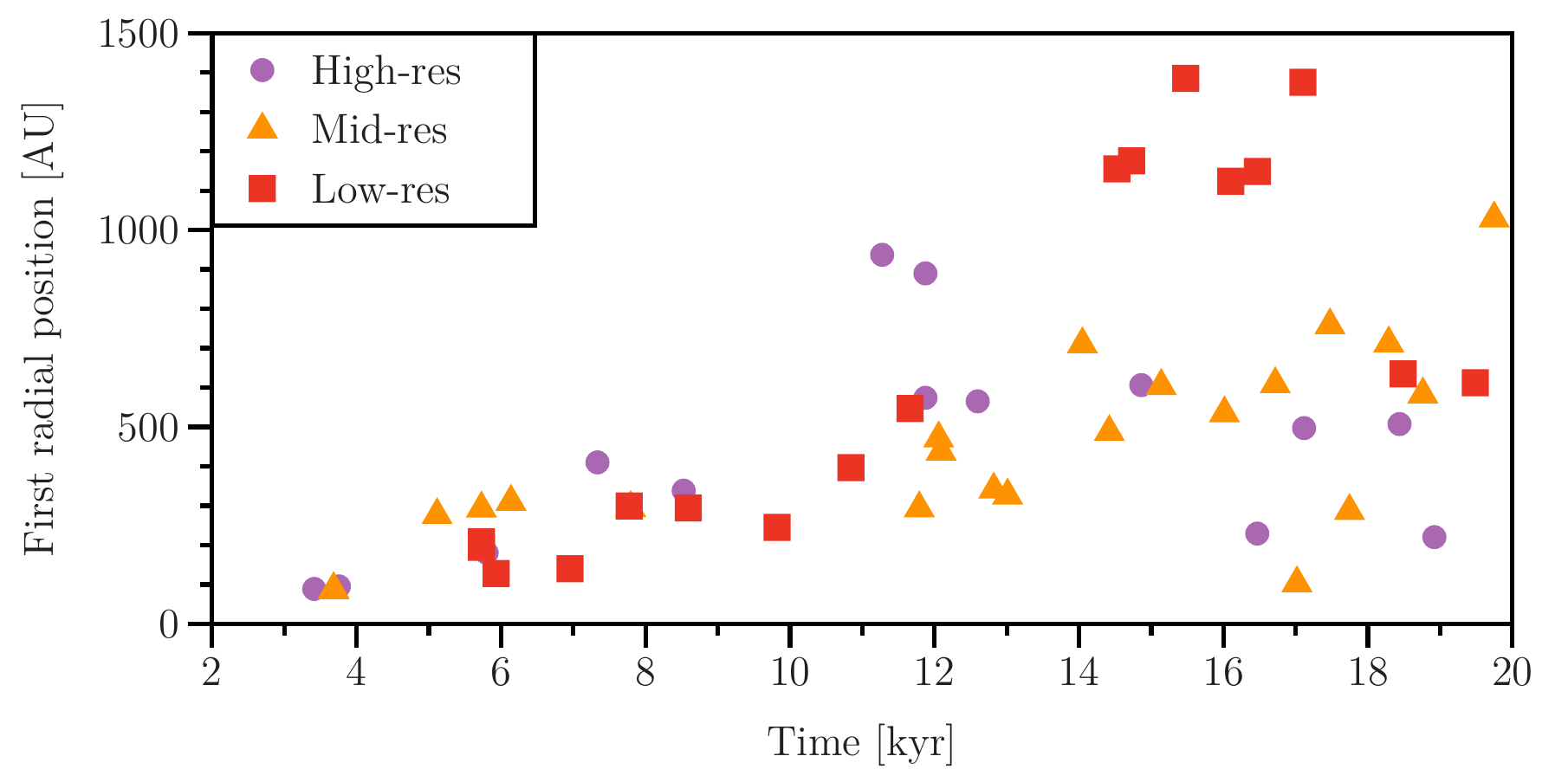} 
    \caption{From top to bottom: number of fragments, total fragment mass, fragment temperature, radius of formation for various resolutions. Only fragments with a lifetime longer than $200$~yr are shown.}
    \label{fig:frag_cvg}
\end{figure}

{From top to bottom, Fig.~\ref{fig:frag_cvg} shows the number of fragments, total fragment mass, fragment temperature, radius of formation in runs  {\tt Low-res} (red), {\tt Mid-res} (orange), and {\tt High-res} (violet).
The number of fragments is between $0$ and $3$ in run {\tt High-res}, between $0$ and $4$ in run {\tt Mid-res} and between $0$ and $3$ in run {\tt Low-res}, showing a good overall agreement.
The total fragment mass shows the same increase trend in runs {\tt Mid-res} and {\tt High-res} (also reported in the main text, in \pluto{} runs), with a similar drop between $10$ and $11$~kyr, which is not the case in run {\tt Low-res}, suggesting again that there is not enough gravitational energy to tap in during an accretion event to prevent further disk fragmentation.
The fragment temperature is slightly smaller in run {\tt Mid-res} than in {\tt High-res}, and even smaller in run {\tt Low-res}, in agreement with the adiabatic contraction heating mechanism.
However, fragments formed in {\tt Mid-res} can still reach the $\mathrm{H_2}$ dissociation limit temperature and exhibit a temperature spike, as in run {\tt High-res}, while it is not the case in run {\tt Low-res}; this suggests that the $10$~AU resolution in {\tt Low-res} is insufficient, for this particular setup - in comparison with the behaviour observed in runs {\tt Mid-res} and  {\tt High-res}.
As shown on the bottom panel of Fig.~\ref{fig:frag_cvg}, the radius at which fragments form can increase with time, as the disk size increases - as shown in the main manuscript - although fragments can still form as well in the innermost parts of the disk.
The new fragments detected at more than $1000$~AU in run {\tt Low-res} come from the close interaction between two fragments and their spiral arms which periodically collide, then fade away and collide again, thus producing a nearly periodic change in the number of fragments between $14$~kyr and $17$~kyr and similar radii of new fragments formation.}

{The fragments properties are qualitatively consistent between runs {\tt Mid-res} and {\tt High-res}, although the number of fragments is slightly higher in run {\tt Mid-res}, suggesting that a resolution of $5$~AU is sufficient to get a consistent qualitative picture (no convergence can be fully achieved since other mechanisms will eventually take place on smaller scales, be it disk turbulence or star/disk interaction), in this particular setup.}

\section{Computational cost and carbon footprint estimate}
\label{app:cost}

Table~\ref{table:cost} gives the computational cost of the simulation presented in the main text and of the lower-resolution simulations presented in Appendix~\ref{app:cvg}. The total is a lower limit since others runs have been performed to test various hypotheses (see e.g. Sec.~\ref{sec:ok20disk}).
As the AMR grid refines regions of interest, in particular around stellar companions, the cost does not strictly scale with the resolution, as could be expected.
Simulations have been performed over $64$ CPU cores.

The $\mathrm{CO_{2,e}}$ ($\mathrm{CO_{2}}$ equivalent) carbon footprint has been computed using the estimate of $4.68\, \mathrm{g/hCPU}$ \citep{berthoud_estimation_2020}.

\begin{table}
\caption{Computational cost (in CPUkhr) and $\mathrm{CO_{2,e}}$ footprint estimate (in kg) of the simulations presented in Sec.~\ref{app:cvg}.}
\label{table:cost}
\centering 
\begin{tabular}{c | c c} 
	\hline\hline
	 Model 				& Cost [CPUkhr] 	&   $\mathrm{CO_2}$ emission [kg] \\ \hline 
	{\tt High-res}			& $20$ 	&	$93.6$  \\ \hline
	{\tt Mid-res}			& $6.9$ 	&	$32.3$  \\ \hline
	{\tt Low-res}			& $4$ 	&	$18.7$  \\ \hline
	Total					& $30.9$ 	& $144.6$ \\ \hline\hline
\end{tabular}
\end{table}

\end{appendix}

\end{document}